\documentclass[final]{agujournalarxiv}

\journalname{JGR-Planets}
\usepackage{graphicx}
\usepackage{url}

\setkeys{Gin}{draft=false}

\begin{document}

%\begin{frontmatter}

\title{Texture and composition of Titan's equatorial sand seas inferred from Cassini SAR data: Implications for aeolian transport and dune morphodynamics}

\authors{A. Lucas\affil{1}, S. Rodriguez\affil{1}, F. Lemonnier\affil{2}, A. Le Gall\affil{3}, S. MacKenzie\affil{4}, C. Ferrari\affil{1}, P. Paillou\affil{5}, C. Narteau\affil{1} }

\affiliation{1}{Universit\'e de Paris, Institut de physique du globe de Paris, CNRS,  Paris, France}
\affiliation{2}{LMD,UPMC, CNRS, France}
\affiliation{3}{LATMOS/IPSL, UVSQ Universit\'e Paris-Saclay, CNRS, Guyancourt, France}
\affiliation{4}{Department of Physics, University of Idaho, Moscow, ID, USA}
\affiliation{5}{Observatoire de Bordeaux, Universit\'e de Bordeaux, CNRS, Bordeaux, France}

\correspondingauthor{A. Lucas}{lucas@ipgp.fr}

\begin{keypoints}
	\item  Physical properties over Titan sand seas are quantitatively derived from Bayesian inference of the Cassini SAR amplitude data

	\item Interdunes terrains are made of fine and coarse parsed grains over a polluted water-ice bedrock

	\item Analogy with terrestrial sand seas in Niger and China suggests  that dunes on Titan reflect current climatic conditions
\end{keypoints}

\begin{abstract}

The texture, composition, and morphology of dunes observed in the equatorial regions of Titan may reflect present and/or past climatic conditions. Determining the physio-chemical properties and the morphodynamics of Titan's dunes is therefore essential to understanding of the climatic and geological history of the largest moon of Saturn. We quantitatively derived average surface properties of dune and interdune areas (texture, composition) from modeling of the microwave backscattered signal and Monte-Carlo inversion of the despeckled Cassini/SAR data over Titan's three largest sand seas: Belet, Shangri-La and Fensal. We present the first analysis of the backscatter functions extracted from despeckled SAR images that cover such a large range in incidence angles, including data from the beginning of the Cassini mission up to its Grand Finale. We show that dunes and interdunes have significantly different physical properties. Dunes are found to be more microwave absorbent than interdunes. Additionally, potential secondary bedforms, such as ripples and avalanches, may have been detected, providing potential evidence for currently active dunes and sediment transport. Our modelling shows that the interdunes have multi-scale roughnesses with higher dielectric constants than the dunes which have a low dielectric constant consistent  with organic sand. The radar brightness of the interdunes can be explained by the presence of a shallow layer of significantly larger  organic grains, possibly non-mobilized by the winds. {Together, our} findings suggest that Titan's sand seas evolve under the current multi-directional wind regimes with dunes that elongate with their crests aligned in the residual drift direction.

\end{abstract}

\section{Introduction}

After 13 years of exploring and revealing the Saturn's system, in particular its largest moon Titan, the Cassini spacecraft plunged \replaced{on Saturn}{into Saturn's} upper atmosphere on Friday 15th of September 2017. Among the most prominent {of} Titan's landforms discovered by the Cassini-Huygens mission are the large sand seas that cover 15-20$\%$ of the global surface of the moon, {located} mostly in the equatorial regions \citep{Lorenz2006,Radebaugh2008,Radebaugh2013,LeGall2011,Rodriguez2014,AharonsonBook2014}. As inferred from the {spectral} observations of the VIMS (Visual and Infrared Mapping Spectrometer) instrument \citep{Soderblom2007,Barnes2008, Rodriguez2014} and \deleted{supported by } the surface properties derived from the radiometry mode of the RADAR instrument \citep{LeGall2011}, these vast sand seas  constitute a major reservoir of organics at the surface.  Consequently, they play a prime role in the geological and climatic history of Titan \citep{AharonsonBook2014}.

\begin{figure*}[!ht]
	\centering
	\includegraphics[width=0.95\textwidth]{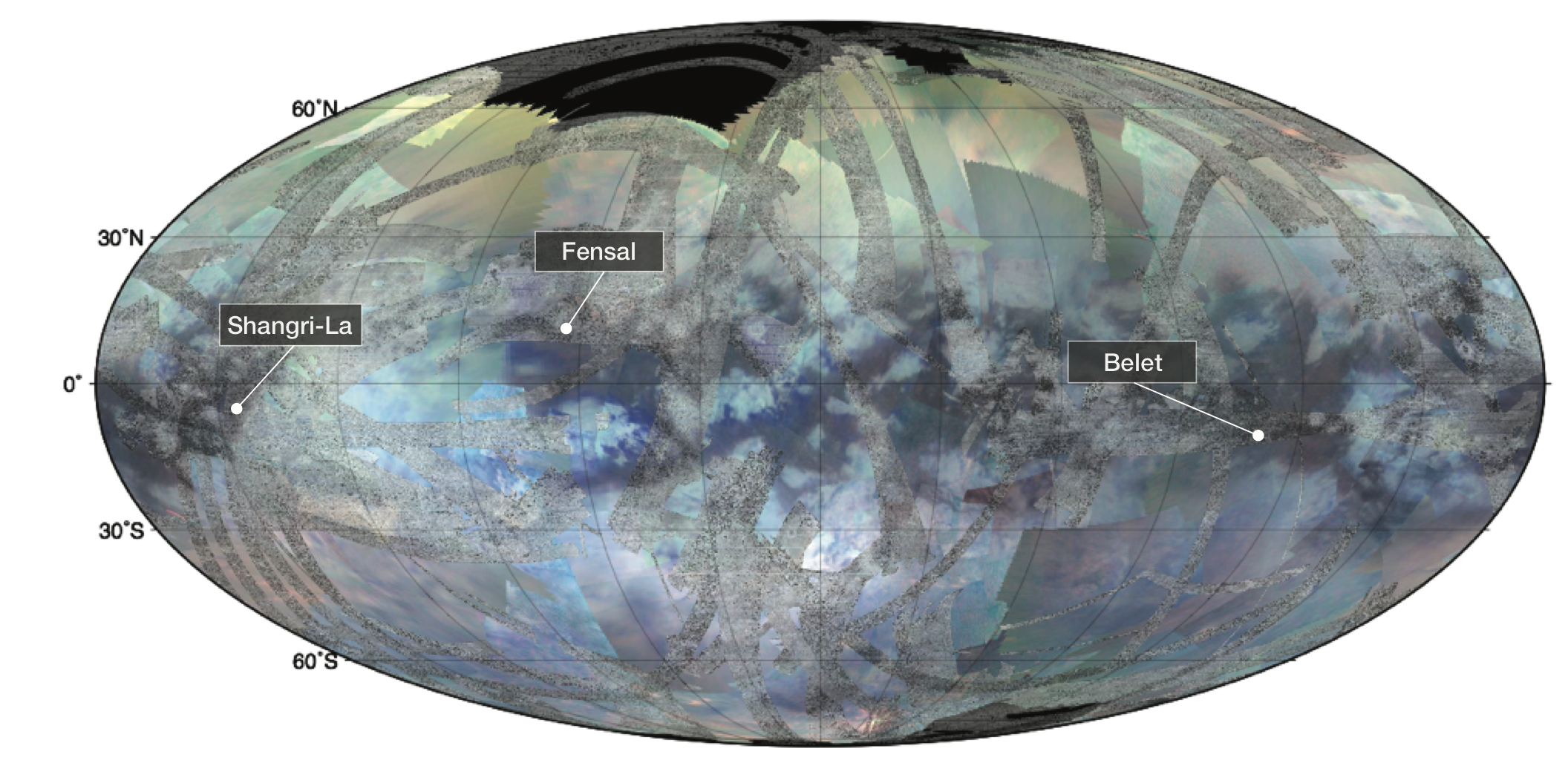}
	\caption{A global map of Titan  showing the location{s} of the regions of interest{:} Shangri-La, Fensal and Belet sand seas. The basemap is an empirically corrected VIMS mosaic from \citep{Rodriguez2014}. SAR swaths are overlain in transparency.  The Hammer projection is centered at 0$^\circ$ longitude with graticules every 30$^\circ$ in both longitude and latitude.}
	\label{fig:mapall}
\end{figure*}

\begin{figure}[!ht]
	\centering
	\includegraphics[width=0.48\textwidth]{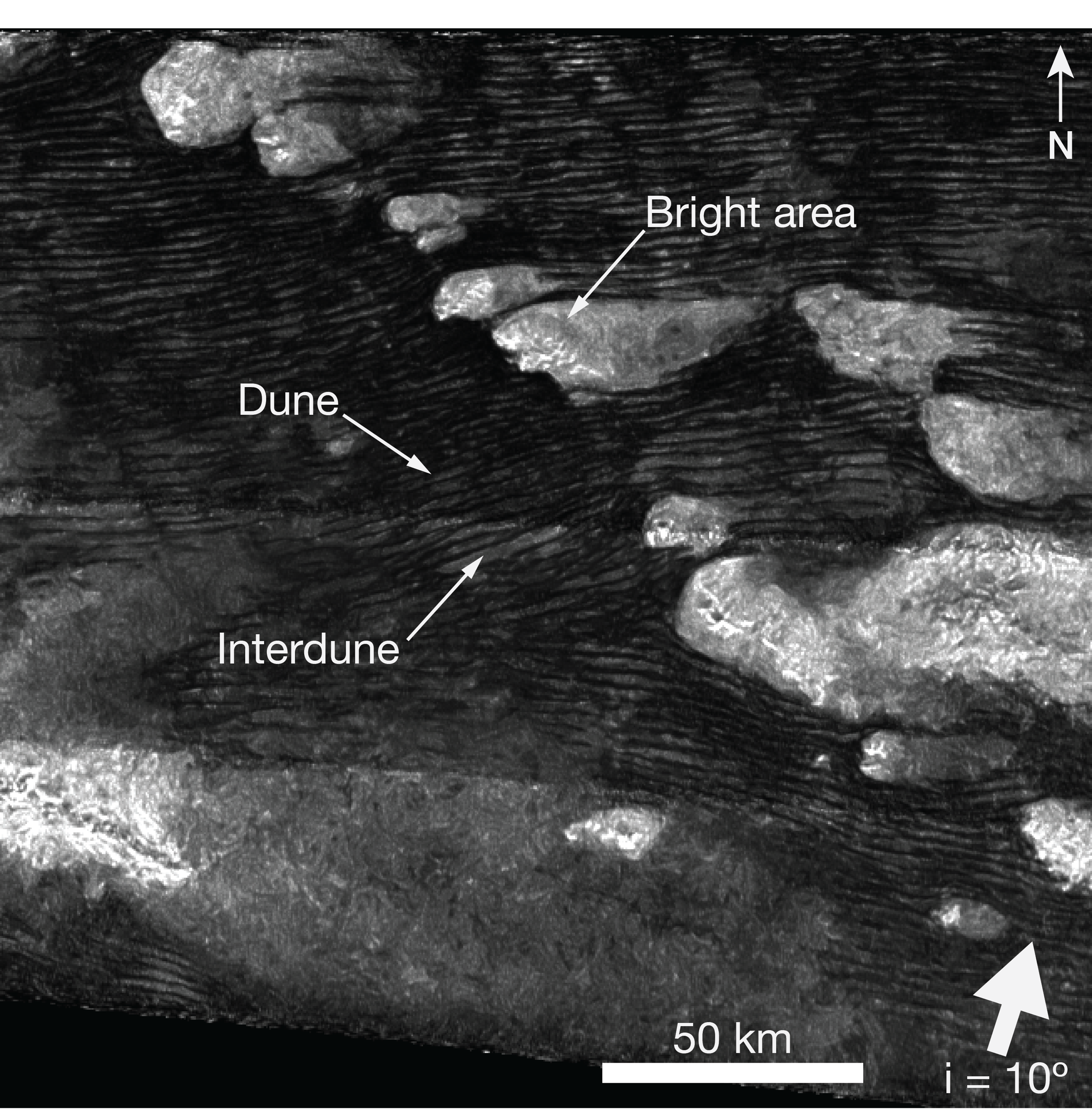}
	\caption{Belet sand sea as observed by the RADAR operating in SAR mode on swath T61 {after} denoising (see main text section 2.2 and \cite{Lucas2014b}). Individual dunes are distinguishable from interdunes. Note that the look direction is roughly normal to the dune's crest orientation (thick white arrow). "Inselbergs" are radar-bright terrains commonly found embedded in Titan's dune fields as obstacles.}
	\label{fig:Belet}
\end{figure}

Thanks to the high spatial resolution of the 5 beam Synthetic Aperture Radar (SAR) imager mode of the RADAR instrument{--} operating at 2.17 cm wavelength (Ku-band) in HH polarization \citep{Elachi2004}{--}individual linear dunes have been identified within those sand seas and mapped at the global scale \citep{Lorenz2006, Radebaugh2008, LeGall2011, Rodriguez2014}. These sand seas present a dark signature {in} visible (ISS camera) and {microwave} (RADAR instrument) data, while they are {``}brown\replaced{ish}{''} {in} hyperspectral data (VIMS) {(i.e. dark at 1.3, 2 and 5$\mu$m)} (Figure \ref{fig:mapall}). Geomorphic analyses show that most of the bedforms are 100-to-1000's km long linear dunes with a crest-to-crest distance around 3 km \citep{Lorenz2006,Lorenz2010,LeGall2011,Savage2014,Lucas2014b}. {Dune heights have been measured as 50-150m high with altimetry and radarclinometry} \citep{Barnes2008, Neish2010, Mastrogiuseppe2014}. In some SAR swaths, interdune areas are clearly distinguishable from the dunes as exemplified in the Belet sand sea shown in Figure  \ref{fig:Belet}. This have been confirmed from hyperspectral observations \citep{Barnes2008,Bonnefoy2016}. Sources for the organic sand have yet to be identified at the surface and {may have even} disappeared as these aeolian systems are thought to be old \citep{Rodriguez2014, Barnes2015, Brossier2018}.  Additionally, radar-bright topographic obstacles called "inselbergs" are embedded in the sand seas. These features are probably outcrops of the bedrock and thus likely have a texture and composition distinct from that of the dunes. Moreover, their properties may vary across Titan: a "radar-bright unit" is considered  as geomorphic group rather than a single geological unit.

As on Earth \citep{Gao2015,Lucas2015} and Mars \citep{Fernandez2018}, assessing the surface properties {of} the sand seas, especially of the dunes and their interdune areas {on Titan}, is  critical because distinguishing between dune origin and nature{--that is, their growth mechanisms}{--} will depend upon {dune and interdune properties (composition, grain size and  thickness of the sediment cover)} \citep{Gao2016,Ping2017}. Moreover, evaluating the properties of sand {sea} "inselbergs" is of importance {because} these geological units could be a proxy for the dune substratum that may {outcrop} in the interdunes areas. Without a more accurate knowledge of the sediment properties (and availability for saltation) in the interdunes and in the dunes, the interpretations of the observed dune shape and orientation in terms of wind regime and dune activity remain ambiguous. Previous studies based on the infrared VIMS observations \citep{Barnes2008, Rodriguez2014,Bonnefoy2016} and microwave radiometry \citep{LeGall2011,LeGall2012,LeGall2014,Bonnefoy2016} have revealed a possible difference in composition and/or grain size in the interdunes compared to the dunes.  {However, {a} statistical inversion of the data with a physical model has yet {to be completed}. }

In this study, we aim to assess Titan's surface and subsurface properties within the different equatorial sand seas based on their Cassini RADAR SAR signature. By combining state-of-the-art SAR image despeckling, filtering and classification techniques, modeling of the microwave backscatter, and taking advantage of the large coverage in incidence angles over the sand seas now available, we seek to determine the bulk surface properties of the dunes,  interdunes and inselbergs at three major sand seas:  Belet, Shangri-La and Fensal (Figure \ref{fig:mapall}). We present the geomorphic classification and microwave backscatter extraction workflow for the SAR data in section 2. We describe our modeling approach and Bayesian inference in section 3. We discuss our results and their implications in section 4 and then {present our conclusions}.

\section{Terrain  segmentation  and backscattering properties from SAR data: Example from Belet}

In this section we describe the data reduction workflow to  separately extract the backscatter cross-section ($\sigma^o$) from the dunes, interdunes and the inselbergs. We  detail below the complete workflow for the Belet sand sea as an example; the same analysis has been applied to the Shangri-La and Fensal sand seas (see the Supporting Information).

\subsection{Sand seas as seen by the RADAR SAR imager}

\begin{figure}[!htp]
		\centering
\includegraphics[width=0.56\textwidth]{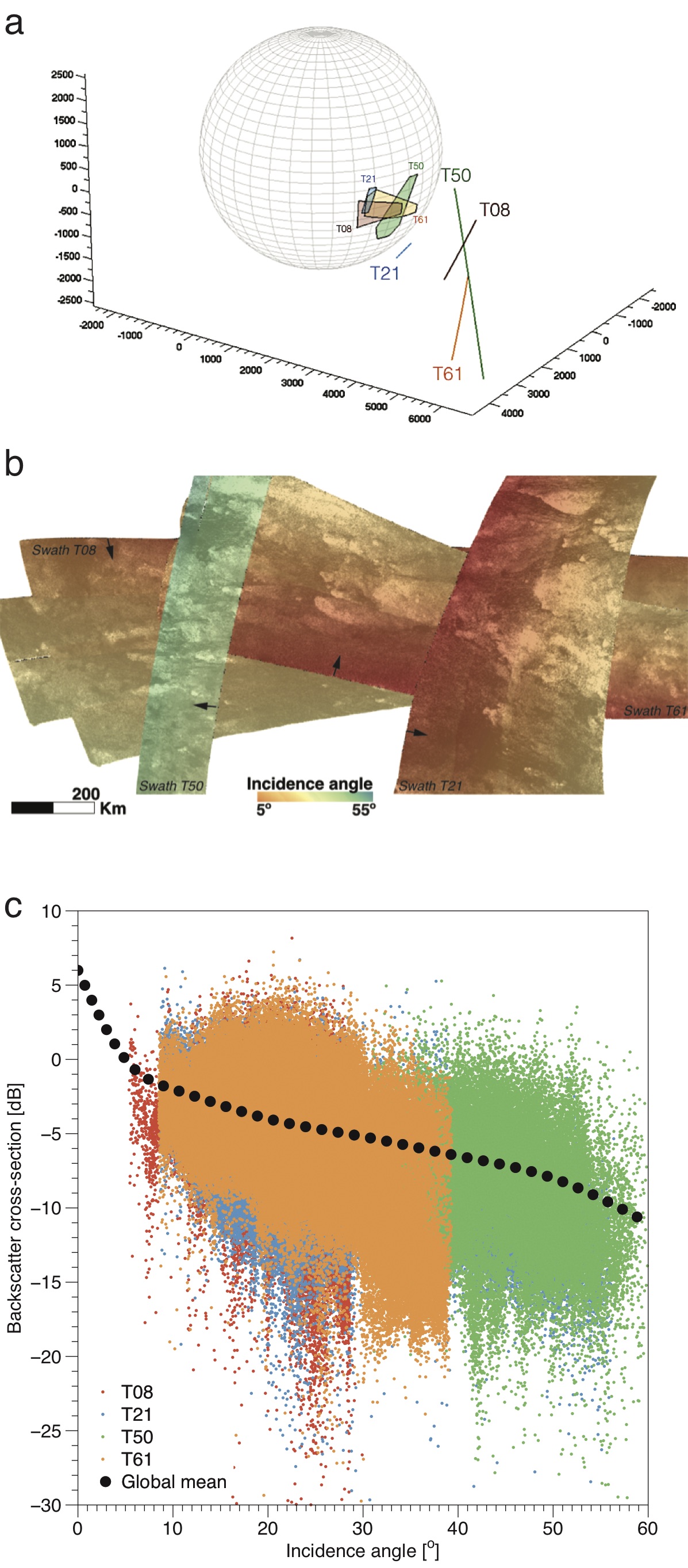}
	\caption{a. Geometries of RADAR observations: footprints and associated Cassini orbits for T08, T21, T50 and T61 swaths over the Belet sand sea. b. Amplitude backscatter and observational incidence angle for the 4 SAR observations.  Color scales with nominal incidence angle. Black arrows show look direction for each swath. c. Backscatter cross-section (expressed in dB) as a function of incidence angle for each individual swath compared to the global mean computed {across} all SAR observations after Cassini.}
	\label{fig:geom}
\end{figure}

In SAR mode, the RADAR experiment provides the highest spatial resolution available of Titan's surface (down to 300 m), allowing the detection of individual dunes as shown in Figure \ref{fig:Belet}. In addition, some regions have now been observed with a wide range of acquisition geometries in between non-circular Cassini flybys (see Figure \ref{fig:geom}). In particular, Belet has been imaged with incidence angles ranging between 8$^\circ$ to 55$^\circ$  and \deleted{almost in the four main cardinal directions (i.e.,} with a wide range of look directions\deleted{)} (Figure \ref{fig:geom}) thanks to the partial overlapping of seven RADAR swaths from Titan's flybys{: T08, T21, T50, T61, T84, T91 and T92. For the sake of clarity, only the first four are shown in Figure \ref{fig:geom}.}
Note that Shangri-La and Fensal have been similarly observed (see Appendix A1 and the Supporting Information for the summary of Titan's observations used in this study). No additional data will be available until the next mission to Titan.

Such a  wide range of observation geometries offers a unique opportunity to determine the microwave backscatter behavior (i.e., $\sigma^o$ as a function of the incidence angle) and therefore the textural and compositional properties of different terrains over the three sand seas {considered}. We note that a similar breadth of geometries does not exist for Earth's deserts. Thus, the scope of our comparison between this work and terrestrial studies in section 4 is limited {structural rather than radiometric analogies} (shape and orientations of the dunes, grain size sorting between dunes and interdunes, sediment availability in the interdunes{, etc.}).

\subsection{Data reduction and classification}

\begin{figure}[!ht]
	\centering
	\includegraphics[width=0.66\textwidth]{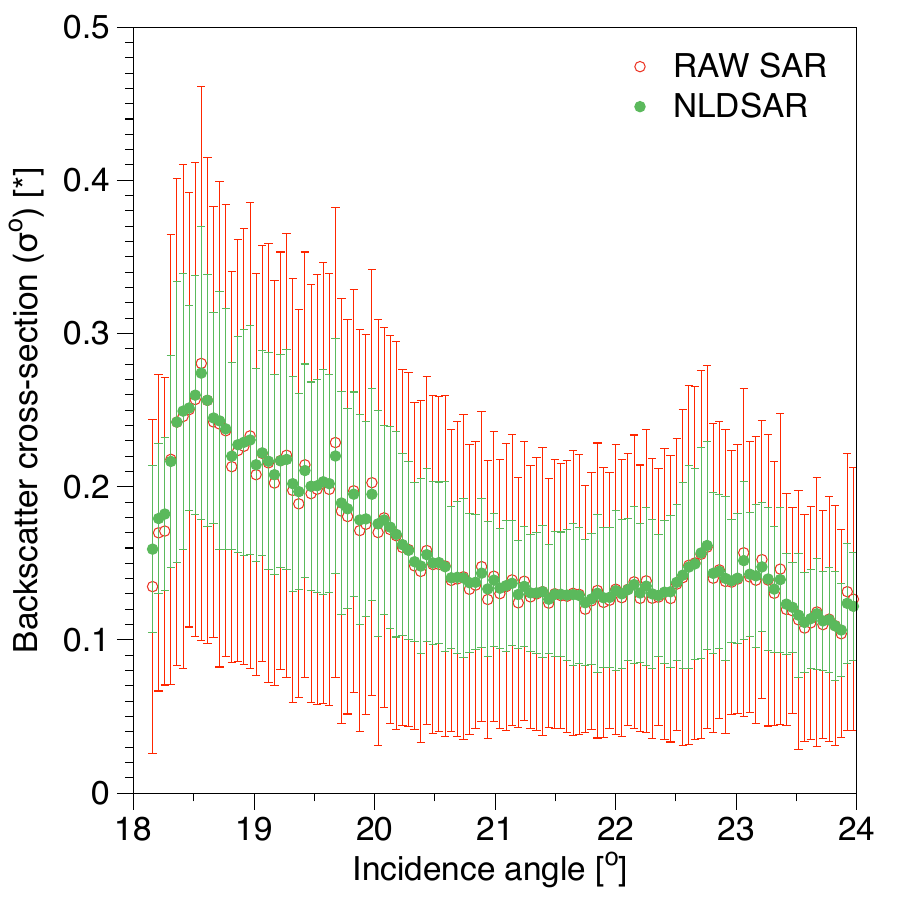}
	\caption{Backscattering cross-section $\sigma^o$ in physical unit (linear scale) as a function of the incidence angle over Belet as observed on T08 swath for raw SAR (red) and NLDSAR (green) data sets. Dispersion is given for 1 standard deviation.}
	\label{fig:rawvsnlm}
\end{figure}

Cassini SAR data suffer from speckle noise, hindering the detection of fine details and quantitative analysis. We therefore use the despeckled SAR data,  NLDSAR {(}Non-Local Denoised SAR) in our work. The efficiency and conservative properties of this non-local denoising is discussed {in detail} in {\cite{Lucas2014b}}. While significantly improving the SAR image quality and sharpness, {which is} of great use for pattern identification and detection, we show that the denoising process  {does not alter} the physical values of the backscattering cross-section ($\sigma^0$). This is essential when using the data for the inversion with a physical model. \replaced{As exemplified in Figure \ref{fig:rawvsnlm}, where the backscatter function over the Belet sand sea (extracted from the T08 swath) is shown, we greatly reduce the data dispersion thanks to the use of the NLDSAR image while preserving its mean values.}{As shown in Figure \ref{fig:rawvsnlm}, we greatly reduce the data dispersion in the backscatter function over the Belet sand sea while preserving the mean by using NLDSAR rather than the raw SAR.}

\begin{figure}[htp]
	\centering
	\includegraphics[width=0.5\textwidth]{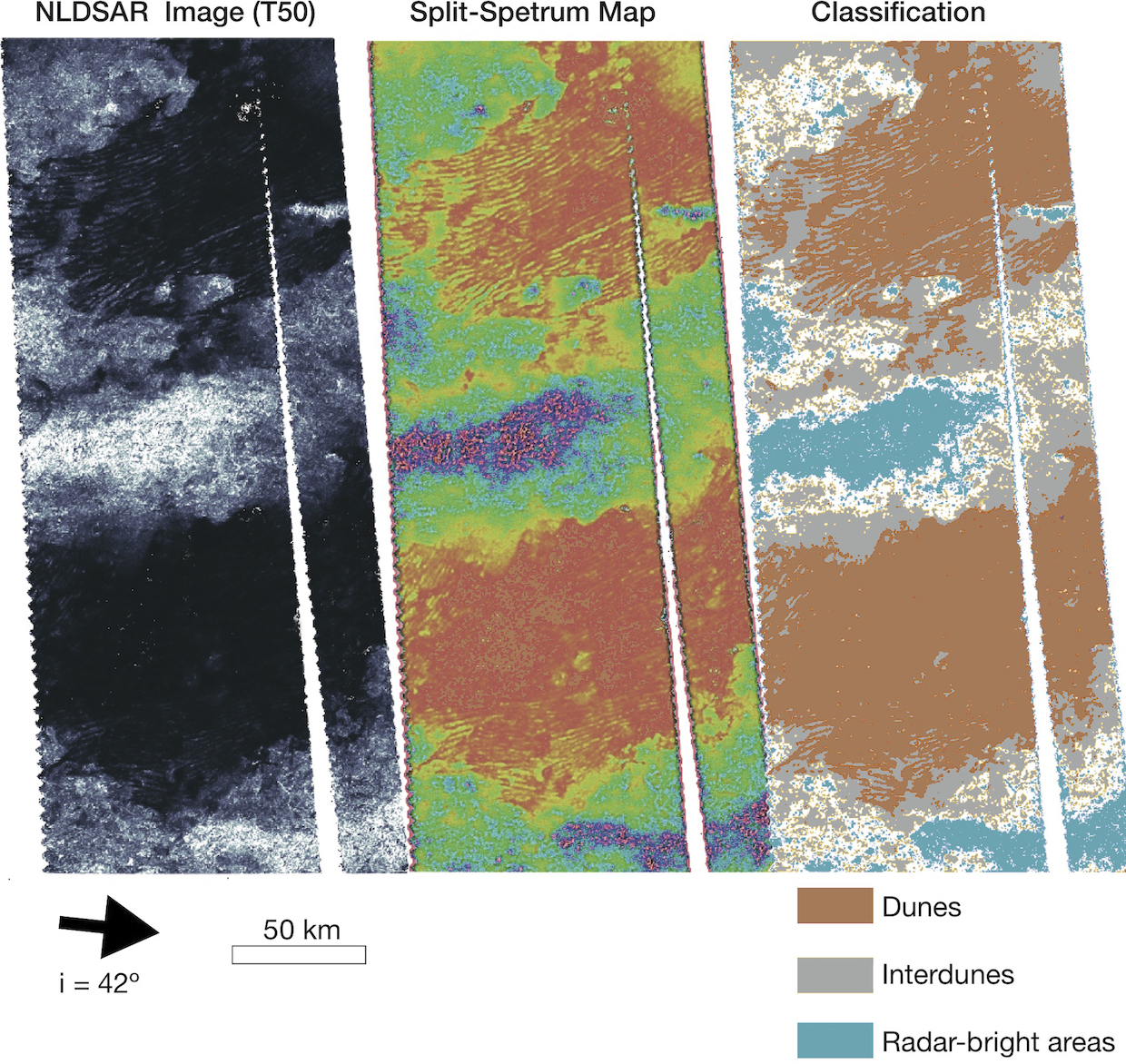}
	\caption{Geomorphic classification {after} denoising and split-spectrum analysis exemplified over Belet sand sea from the T50 NLDSAR swath. Dunes are shown in brown. Interdunes in gray are detectable even over the dark areas. Radar bright terrain (e.g., inselbergs) are shown in light blue.  Blank areas correspond to non-classified terrain. {This analysis has been conducted} on all swaths covering the studied areas and are presented in the Supplementary Information.}
	\label{fig:classif}
\end{figure}

\begin{figure}[htp]
	\centering
	\includegraphics[width=0.5\textwidth]{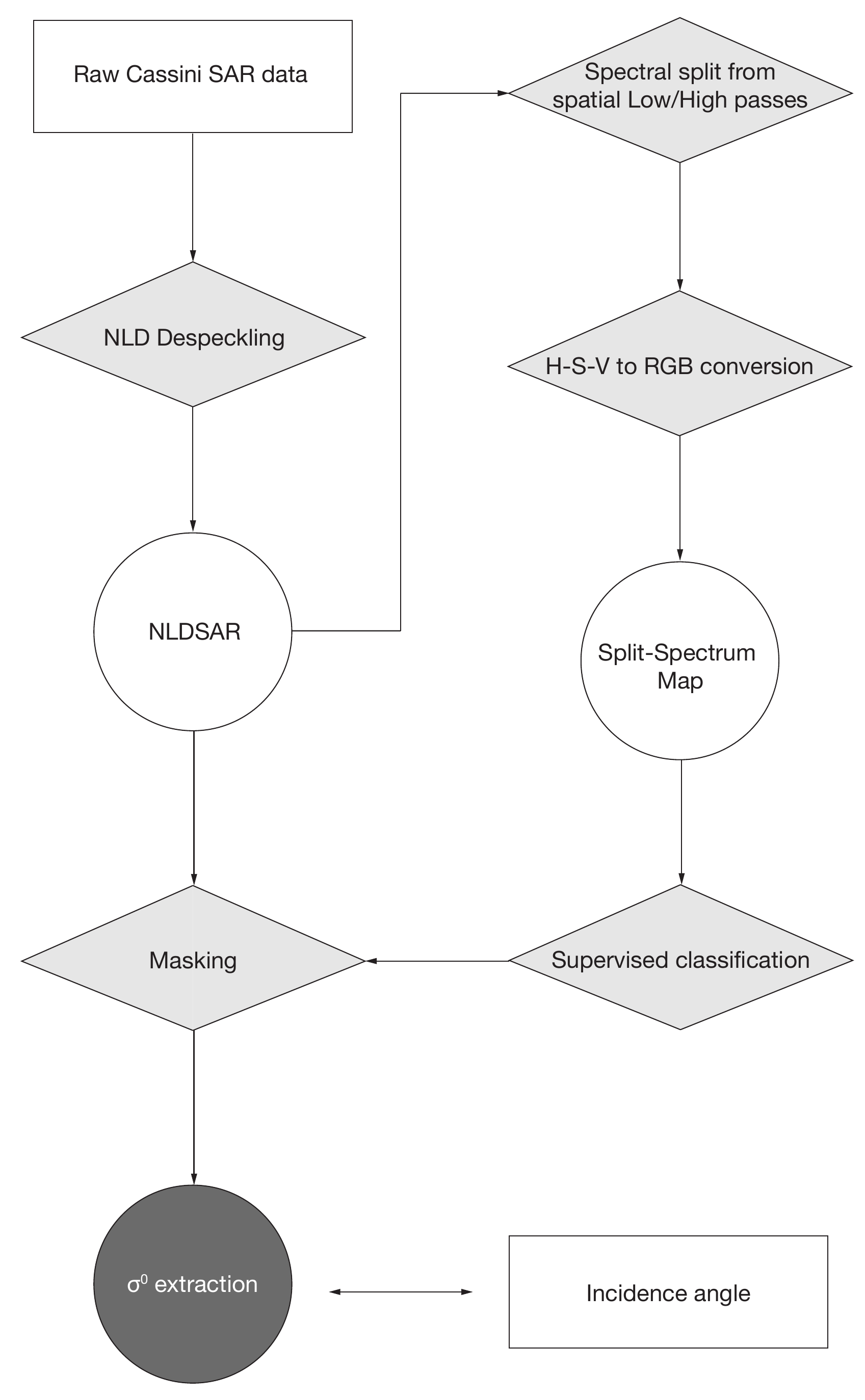}
	\caption{Workflow steps for extracting backscatter signal ($\sigma^0$) from despeckling, split-spectrum and supervised classification.}
	\label{fig:Workflow}
\end{figure}

Taking advantage of the NLDSAR data, we {then} develop a complete workflow in order to automatically extract the backscatter functions of dunes, interdunes and inselbergs. Geomorphic unit segmentation is performed by a split-spectrum analysis over the SAR data inspired by \cite{Daily1983}. As discussed {in} that previous study, low spatial frequency content is due to backscattering signal variations over the surface and subsurface while high frequency content is associated with local topographic features. Low-pass and high-pass filters are hence applied to the SAR data in order to separate low and high spatial frequency information. Additionally, we define a fixed saturation threshold that controls the resulting brightness of the split-spectrum map. This parameter plays a negligible role in the geomorphic unit extraction hereafter. The low frequency content can {then} be assigned to the Hue axis, the high frequency content to the Value axis, and the fixed saturation threshold to the Saturation level, resulting in a colored image defined in the Hue-Saturation-Value colorimetric domain. Finally, we convert the H-S-V values into the R-G-B domain to create a synthetic-colored image based on the spatial frequency content of the $\sigma^o$ amplitude, hereafter referred to as the split-spectrum map. The split-spectrum map created from NLDSAR data greatly facilitates classification, as shown in Figure \ref{fig:classif}. Finally, the backscattering cross-section values as {a} function of incidence angle are extracted from the NLDSAR data upon the split-spectrum segmentation.  The complete data reduction workflow, including the main steps, is summarized on Figure \ref{fig:Workflow} and has been applied to observations of Belet, Shangri-La and Fensal sand seas (Figure \ref{fig:mapall}, Figure \ref{tab:data}, and Supplementary Information).

\subsection{RADAR backscattering cross-section as a function of the incidence angle}

 We computed the mean within a 3-sigma dispersion of the backscatter cross-section every half-degree of incidence angle. The mean and dispersion are calculated over thousands of measurements for each considered incidence angle, guaranteeing the statistical reliability of the extracted backscatter functions. This is done for each of the 3 geomorphic units (dunes, interdunes and inselbergs) isolated over the Belet sand sea.

Note that the incidence angle of observation is not necessarily the local incidence angle{,} especially for the dunes and the inselbergs {which} are not flat surfaces. {W}e expect that the dunes {have} lee sides with avalanches {close} to the angle of repose (30$^{\circ}$) \citep{Neish2010} based on glints observed in T8 and T61 (Supplementary Information), and that the bright-areas might present local slopes as steep as 75$^{\circ}$ \citep{Tomasko2005}. However, the Cassini RADAR topography products (SARTopo, altimetry, stereo-derived digital elevation maps) cannot {precisely} resolve local slopes over the sand seas because of limited coverage and/or coarse spatial resolution \citep{Lorenz2013,Corlies2017}. Direct information on the local scale topography is therefore missing. We choose to deal with this issue by averaging the backscattering cross-sections into 0.5-degree incidence bins, and retain the resulting values when more than 10,000 pixels per bin is reached. Thus, surface facets with different azimuths and slopes are integrated and  topographic effects, if any, are balanced. This is {especially true for the dunes}, where limited spatial resolution of the SAR images generally impedes the distinction between their two sides even at the scale of a single SAR pixel. Our present study therefore focuses on the "representative" properties of the dunes, interdunes and inselbergs geomorphic units, averaged over the entire Belet, Shangri-La and Fensal sand seas. Local study at the scale of the dune or interdune objects is {beyond} the scope of this work but will be the subject of future studies.

 \begin{figure}[!ht]
 	\centering
 	Belet \hspace{1in} Shangri-La \hspace{1in}  Fensal
 	\includegraphics[width=.99\columnwidth]{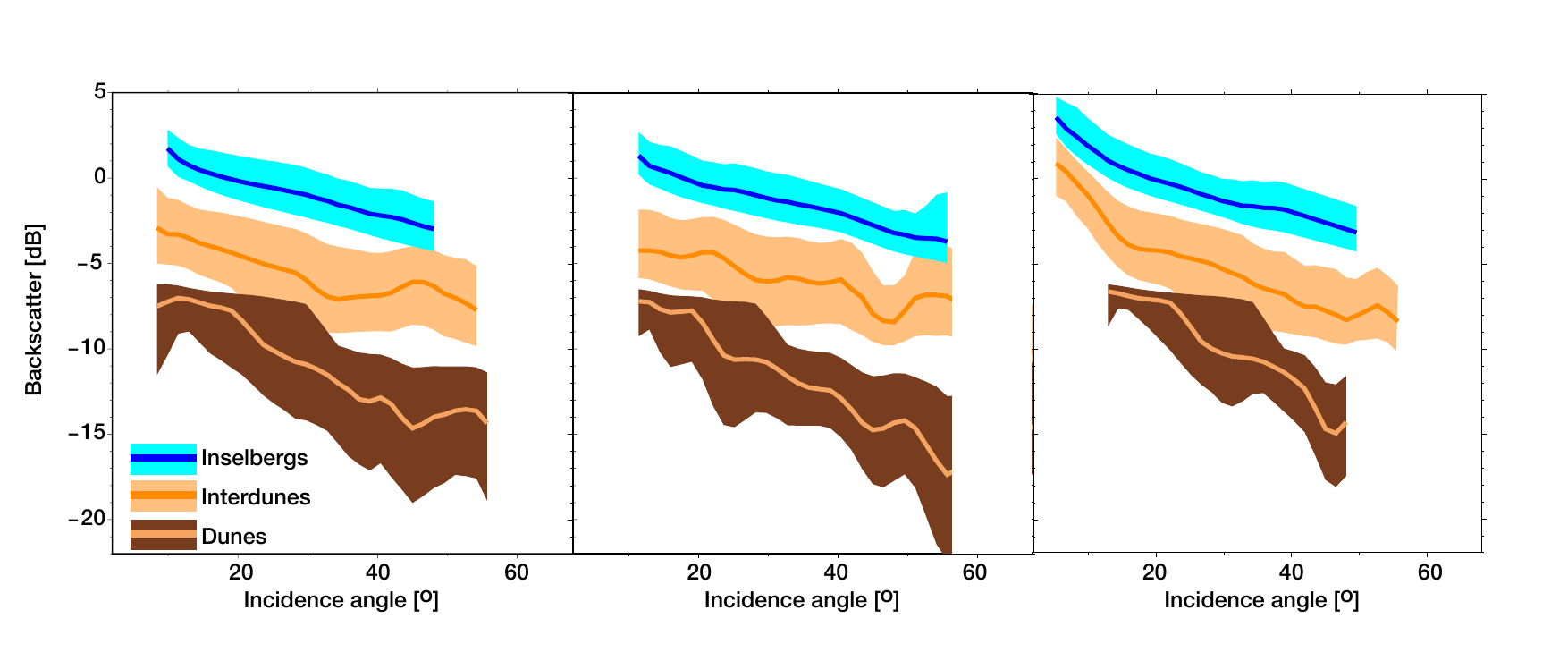}

 	\caption{Backscatter functions over the three sand seas (Belet, Shangri-La and Fensal from left to right) and the three geomorphic unit{s} considered: inselbergs (blue), interdunes (light brown) and dunes (dark brown). The mean value{s} with the 2-sigma dispersion are shown. }
 	\label{fig:sigma0_all}
 \end{figure}

The observed backscattering cross-section as a function of the incidence angle over the three sand seas is shown in Figure \ref{fig:sigma0_all}. We emphasize that these are the first backscatter functions ever  extracted from denoised SAR images that cover such a large range of incidence angles since the beginning of the Cassini mission. Interestingly, the microwave backscatter functions of the 3 geomorphic units do not overlap, suggesting significant differences in their surface and subsurface properties.
The inselbergs show a {gently} decreasing trend with increasing incidence angles:  -0.09 dB/$^\circ$ when accounting only for the pure{ly} non-specular part of the curve ($\theta > 15^\circ$) and -0.11dB/$^\circ$ for the whole curve. This is an indication of a very large roughness with respect to the Cassini RADAR wavelength of 2.17 cm \citep[see][]{Ulaby1982} and/or a strong volume scattering in the subsurface as proposed by \cite{Paillou2006}. Though lower {in} backscatter cross-section, the interdune areas also present a similar trend of -0.09 dB/$^\circ$ {on} average for the three dune fields. The dunes, on the other hand, {have} an even lower value of the {backscatter} function with  a strong{ly} decreasing trend with incidence angle of -0.24 dB/$^\circ$ on the non-specular part and -0.18 dB/$^\circ$ for the whole curve. This {explains} their darker {appearance} compared to the interdunes and inselbergs at all viewing geometries across all {Cassini} SAR observations.  This may reflect a textural difference (i.e.,  lower roughness), a lower relative permittivity and/or a lower diffuse  (i.e., roughness and/or volume scattering) component \citep{Ulaby1982}. The overall flatness of  the curves for the three units between 10$^\circ$ and 50$^\circ$ (e.g., 20$^\circ$ and 50$^\circ$ for the dunes) suggests a significant diffuse component due to roughness or/and volume scattering. These will be further quantified in the following section by the use of microwave backscattering models.

\section{Microwave backscatter modeling and inversion results}

In this section we first discuss the fundamentals of microwave backscatter modeling and then explain our choices for the subsequent inversion of {the} surface and subsurface properties of the considered terrains \replaced{from}{with} Bayesian inference.

\subsection{Microwave backscattering forward models}

Microwave backscatter modeling has been widely used for terrestrial and planetary applications. We base our analysis on previous studies that have developed models that depend on surface roughness and relative permittivity ({represented as} $\epsilon = \epsilon' - j \epsilon"$), accounting for coherent and single-scattering parts of the signal \citep{Fung1992,Fung1994,Ulaby1982,Paillou2006}.

The surface roughness is defined by two empirical statistical descriptors: the root-mean-square height ($\xi$), quantifying the vertical variations of the surface, and the correlation length ($\zeta$), characterizing the typical horizontal size of surface bumps. We usually define surface roughness with a Gaussian height distribution and exponential autocovariance function $\rho_N$:

\begin{equation}
\xi = \sqrt{\frac{\sum_{i=1}^n (z_i -\bar{z})^2}{n-1} },
\end{equation}
$z$ being the elevation with $\bar{z}$ being its mean, $n$ the number of measurements along the profile, and
\begin{equation}
%\rho(x') = \sum_{i=1}^{N+1-j} z_i z_{j+i-1} \times (\sum_{i=1}^N z_i^2)^{-1}
\rho_N (\delta) = \frac{1}{\xi^2} \big < z(x_i + \delta)z(x_i) \big >,
\end{equation}
with $\big < \big >$ being  the averaging operator and $\delta$ being the measured distance  from {a} considered  point $x_i$. The correlation length {is given by}
\begin{equation}
\zeta = \rho_N^{-1}(\exp(-1)).
\end{equation}

{We then choose} the surface backscatter term. {For the range of} very smooth to very rough terrain, 1-layer Integral Equation Method (IEM), PO (Physical Optics)  and GO (Geometrical Optics) surface scattering models are appropriate (see Appendix A2 for {further} details). Here we use the validity domains defined by equation (\ref{eq:sppchoice}) following \cite{Paillou2006} (Appendix Figure \ref{fig:valid}). {I}t is important to recall {a} few limitations of the considered models.

$\bullet$ Note that the {valid domains} for the surface term (see $\sigma^0_{Spp}$ in Appendix A3) differ slightly throughout the literature \citep[e.g.,][]{Paillou2006,Fung1992,Zribi2002}. Moreover, it has been shown that for terrestrial terrains, models are not always able to reproduce ground truth observations in their theoretical domain of validity due to some misconception and/or {poorly} defined surface roughness {field measurements} \citep{Zribi1997,Zribi2000,Zribi2002} (See Appendix A3).

$\bullet$ As {mentioned} above, the non-coherent surface scattering depends on the two statistical descriptors of the surface roughness ($\xi$ and $\zeta$), which, as defined, are bound to one  another (see Appendix A2). Consequently, they cannot be {inverted} independently.

$\bullet$  The ability of these descriptors to characterize natural surfaces is, in practice, not straightforward, as they depend on the scale at which they are defined (i.e., the length of the roughness profile) as shown by {\cite{Baghdadi2000}}. Moreover, the interdependence of the two statistical descriptors is itself a function of this scale and the nature of the terrain (i.e., rough vs. smooth) \citep[e.g.,][]{Baghdadi2000,Zribi2002,Bretar2013}. It has been shown that it is  more relevant to consider the ratio of the two descriptors in order to characterize the roughness of a natural surface{: the }root-mean-square slope ($s = \xi / \zeta$). Note that only the GO model depends directly on the surface slope $s$ and hence provides a better representation of the surface roughness. {Although previous studies suggest that the ratio $Z = \xi^2 / \zeta$ may be a more satisfying descriptor of roughness, it cannot be directly related to the backscattering modeling as described above and hence {is not considered} in our study}.

The Snell-Descartes laws assume {flat interfaces} and that deviations from planarity {do} not exceed a wavelength fraction. A surface is therefore considered to be rough if the mean quadratic deviation of surface irregularities {satisfies} the Rayleigh quality criterion \citep{Ulaby1982}:

\begin{equation}
\xi > \frac{\lambda}{8*\cos(\theta)},
\end{equation}

$\lambda$ being the wavelength and $\theta$ the incidence angle.  Based on the Cassini RADAR wavelength of 2.17 cm, the validity domain for each model (Appendix \ref{fig:valid}), the roughness of natural terrains (Appendix Figure \ref{fig:rough})  and  previous works from \cite{Paillou2014},  all {Titan equatorial} terrains appear rough and fall into the GO model domain of validity. We confirm this by Monte Carlo simulations accounting for 1-layer models with no model parity: >90\%  of our  reduced $\chi^2$-tests best fits  are obtained with the GO model (Appendix \ref{fig:dirflo}). Consequently, subsequent analysis is only done with the GO model for which the {backscattering} function reads:

\begin{equation}
\sigma^{0}_{Shh} = \Gamma(\theta = 0) \frac{\exp(-\tan^2 (\theta) /2m^2)}{2m^2 \cos^4 (\theta)},
\label{eq:go}
\end{equation}

with $m = \sqrt 2 \xi / \zeta$ and $\Gamma$($\theta$ = 0) being the Fresnel reflectivity at normal incidence angle  (i.e., $\theta = 0$). Recently, \cite{Paillou2006,Paillou2014} applied these models to the Cassini SAR data and emphasized i) that Titan's surface is rough, with respect to the RADAR's wavelength confirming our preliminary analysis, ii) the importance of volume scattering from the subsurface. These findings are also supported by scatterometry and radiometry observations \citep{Zebker2009,Janssen2016}. To account for the volume scattering contribution, we also add the following term:

 \begin{equation}
 \sigma^0_{Vhh} = \frac{3}{4} a T^2_{ij} \cos(\theta) \left( 1 - e^{-2\tau\cos^{-1}(\theta_t)}\right),
 \label{eq:vpp}
 \end{equation}
 where $T_{ij}$ is the Fresnel transmission coefficient between two media (i.e., the air and ground), {and} $\theta_t$ is the transmitted incidence angle\replaced{, and}{.} $\tau$ {is} the optical depth defined as $\tau$ = 1/(1-$a$), with $a$ being the microwave albedo of the first interface (i.e, the surface in case of 1-layer model) which ranges between 0 and 1.  Note that the term  $\sigma^0_{Vhh}$ accounts  for the totality of the diffuse part of the backscattering signal. That is, volume scattering and/or surface multiple-scattering processes can not be differentiated.

 The overall modeled backscattering coefficient is then obtained by evaluating $\sigma^{0} = \sigma^0_{Shh} + \sigma^0_{Vhh}$ (in physical units and then converted into dB) {with} varying $\xi$, $\zeta$, $\epsilon$ and $a$.

Although {the Cassini RADAR microwave signal} can penetrate a few 10's of cm beneath the surface,  only 1-layer models are considered in this study \citep{Paillou2006}. The number of unknowns in {the} case of the +2-layer models become{s} high and would require additional constraints on the physical properties of Titan's surface and subsurface that we do not {yet} have. In Titan's case, therefore, it would {provide no} further insights into the physical properties of Titan's geomorphic units \citep{Paillou2006}.

\subsection{Sensitivity analysis over synthetic tests from Bayesian inference}

\begin{figure}[!htbp]
	\centering
\includegraphics[width=1.15\columnwidth]{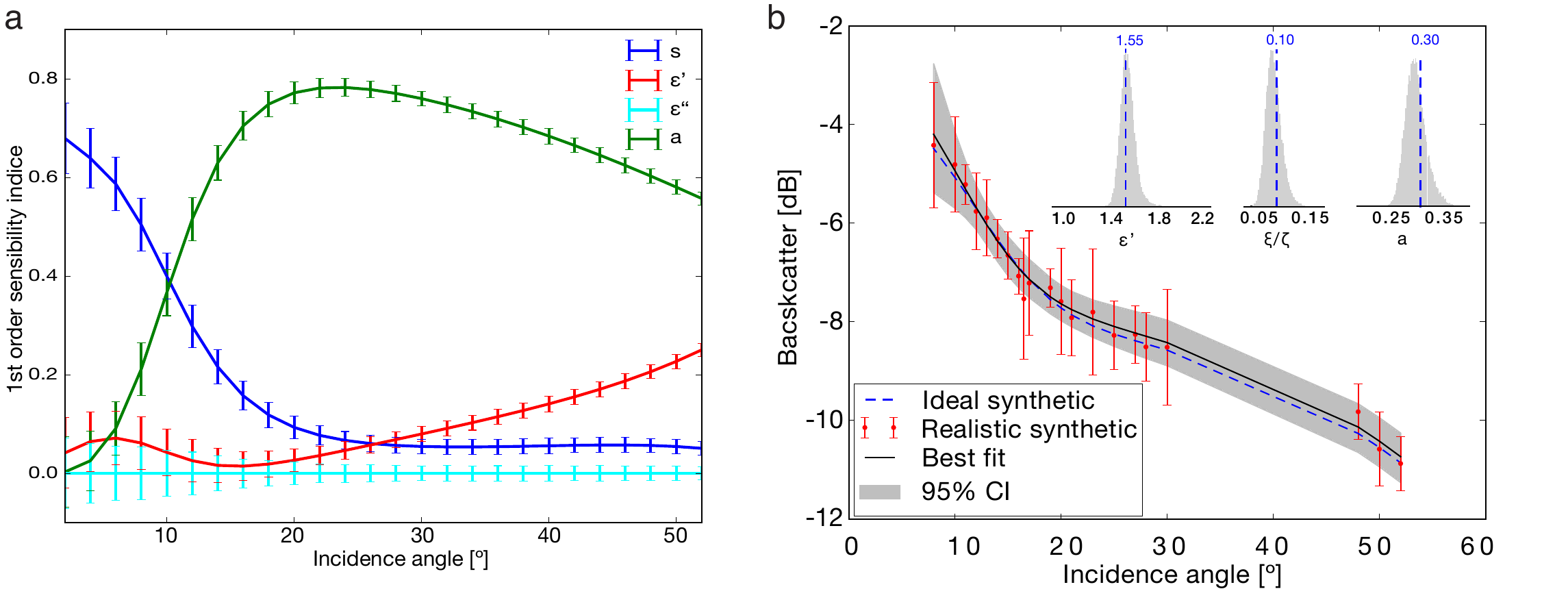}

	\caption{Sensitivity analysis of the microwave backscatter function. (a) Sobol analysis of the GO+Vol. model. At low incidence (i$<$12$^\circ$), the surface roughness through the root-mean-square slope $s$ dominates, while at higher incidence, the albedo $a$, that determines the volume scattering contribution, dominates the  backscattering cross-section. (b) Bayesian inference from the GO+Vol. model on synthetic data. Ideal curve (dashed-blue) is obtained from equations (\ref{eq:go}) and  (\ref{eq:vpp}) with $\epsilon' = 1.55$, $s = \xi/\zeta = 0.10$ and $a = 0.30$. Gaussian noise is added to the mean values and uniform noise is added to the errors. The best fit (black curve) is shown with the associated 95\% confidence interval (gray envelop{e}). \replaced{M}{The m}arginal posterior probability for each parameter is compared to true noise-free values (insert). The respective search ranges are $\epsilon'$ $\in$ [1, 5], $s$ $\in$ [0.005, 0.6] and $a$ $\in$ [0.1, 1.0].}

	\label{fig:synth}
	\label{fig:sobol}
\end{figure}

The non-linearity and non-homogeneity of the modeling make the physical parameter inversion non-trivial. As we aim at being as quantitative as possible, we performed a sensitivity analysis {with} the variance-based Sobol's method which quantifies the amount of variance that each parameter contributes to the unconditional variance of the model output \citep{Sobol2001}. These amounts, caused either by a single parameter or by the interaction of two and more parameters, are expressed as sensitivity indices (SI), which represent fractions of the unconditional model output variance. The first-order SI are the direct variance-based measure of  the contribution to the output variance of each parameters. The case of a very rough surface, including the volume scattering term (i.e., GO model + vol. scat.) is shown on Figure \ref{fig:sobol}a for all physical parameters considered.  Our sensitivity analysis quantitatively demonstrates that the effect of surface RMS slope ($\sqrt{2} s$, with  $s = \xi / \zeta$) dominates at low {incidence angles}, while the contribution of the albedo $a$ dominates at higher angles. Additionally, the contribution of the real part of the relative permittivity $\epsilon'$ remains low ($<$20\%). The GO model {is insensitive} to its imaginary part. However, the low uncertainties {of the first order SI} emphasize that even low-contribution parameters can still be assessed as long as the incidence coverage is sufficient.

In order to compute the marginal posterior probability of each physical parameter, we use a Bayesian statistical model and fitting algorithms based on {the} Markov Chain Monte Carlo inversion (MCMC) method \citep{Salvatier2016}. We assess the capability of the method to correctly retrieve the physical parameters\replaced{, synthetic tests are done}{using synthetic tests} as illustrated in Figure \ref{fig:synth}b. In this example{,} the backscatter function is computed from the GO+Vol. model with  $\epsilon' = 1.55$, $s = \xi/\zeta = 0.10$, and $a = 0.30$ for  a variety of incidence angles randomly drawn with significant gaps between 30$^\circ$ and 50$^\circ$. Gaussian noise is independently added on the mean values for each incidence angle with a standard deviation of 0.3 dB which corresponds to the $\sqrt{N_{RMS}}$ of the derived noise obtained from NLDSAR {technique} \citep{Lucas2014b}. Uncertainties are simulated from a uniform distribution with 1-$\sigma$ at 0.6 dB. We therefore obtain realistic synthetic data that mimic the actual data from Cassini SAR (Figure \ref{fig:synth} and Appendix A3).  As shown in the example in Figure \ref{fig:synth}b, true parameters are accurately assessed by our inversion scheme. This method also presents the advantage {of providing} confidence intervals for each of the retrieved parameters.

\subsection{Titan's equatorial regions microwave properties from Bayesian inference}

\begin{figure*}[!htbp]
	\centering
	\includegraphics[width=1.2\textwidth]{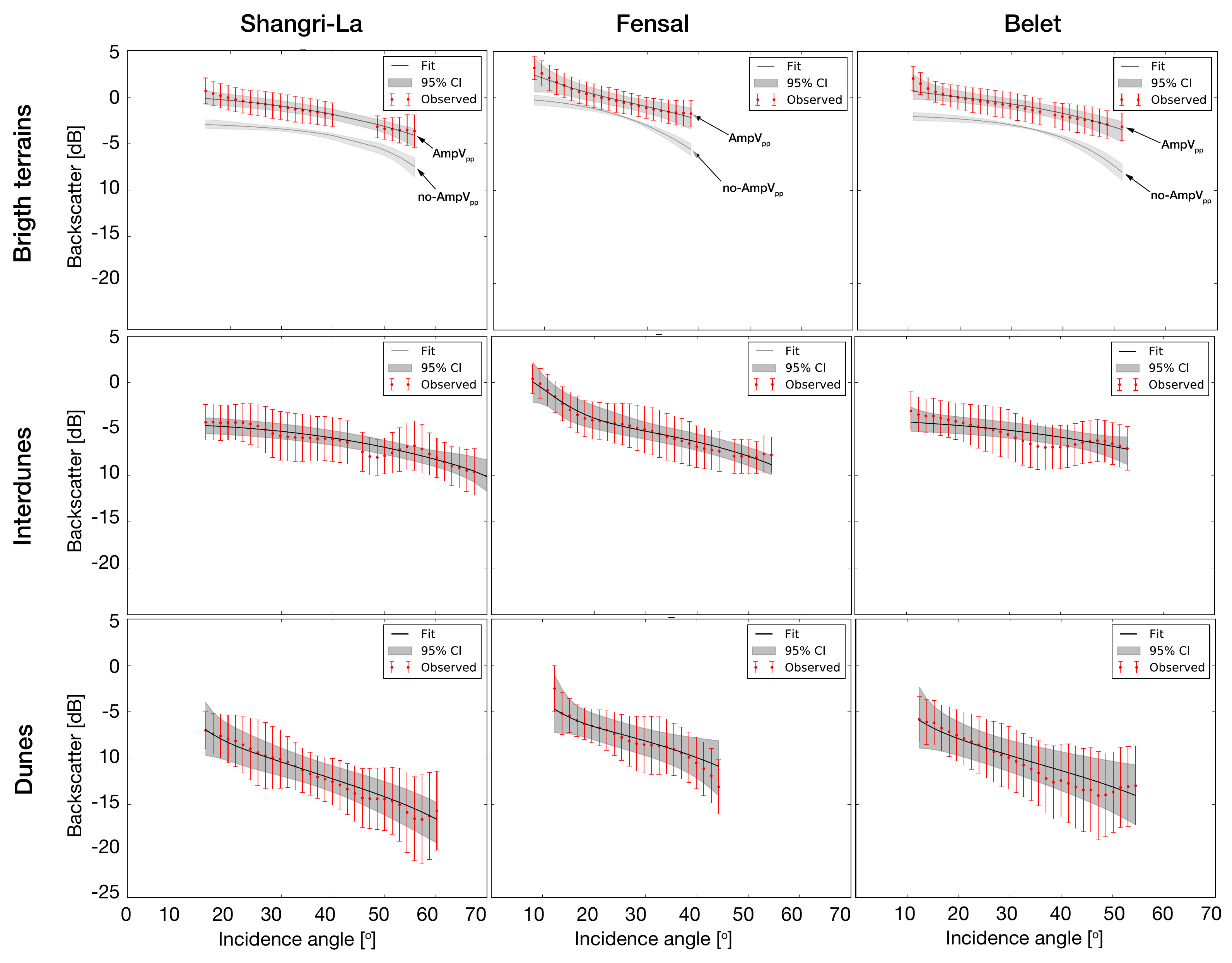}

	\caption{Bayesian inference over Titan's equatorial sand seas.  Backscatter (expressed in dB) as a function of the incidence angle (in degrees) for the dunes, the interdunes and inselbergs (in red) with associated Markov chain Monte Carlo inversion (plain dark curves). The 95\% CI are represented by the gray envelopes. For bright terrains, "Amp" and "no-Amp" indicate results derived with and without accounting for volume scattering amplification discussed in the main text.}
	\label{fig:fit}
	\label{fig:sigma0}
%	\label{fig:probmarg}
\end{figure*}

In order to characterize the physical parameters $s$, $\epsilon$' and $a$ which reflect the surface roughness (i.e., RMS slope), the composition of the terrain ($\epsilon$' may also depends on porosity),  as well as the subsurface contribution respectively, similarly to the synthetic tests, we perform a Bayesian inference over the GO+Vol. model based on advanced Markov Chain Monte Carlo method \citep{Salvatier2016}. This analysis has been conducted with 50{$\times$}10$^6$ runs of the backscatter model for $\epsilon' \in$ [1,5], which corresponds to the expected range of values for material relevant to Titan's surface at 90 K and 13 GHz \citep{Paillou2008}. $s=\xi / \zeta \in$ [0.001,1] {is} based on terrestrial values (see Appendix A1),  and $a \in [0.0,1.0]$. Resulting optimal fits and 95\% confidence intervals (CI) are shown for the three considered units of the three sand seas {in F}igure \ref{fig:fit}. In all cases, the 95\% CI falls into the observed dispersion, even when the incidence coverage is small. The mean observed data are correctly retrieved by our simulations (i.e.,  black lines on Figure \ref{fig:fit}). As discussed below, the synthetic tests support our results on the real observations.  Indeed{,}  these tests consist of both exploring the sensitivity to observation properties (incidence angle) and to the standard deviation and noise function derived from \cite{Lucas2014b}.   Additionally, the spatial distributions  (somewhat randomly distributed) of the physical parameters have been considered in theses tests. As illustrated in Fig. \ref{fig:sobol} and Fig. \ref{fig:synthist}, and Appendix A4, the synthetic tests show that our method does not  breakdown when accounting for the actual data properties (coverage, standard deviation, noise, etc.).

\begin{figure*}[!ht]
	\centering

\includegraphics[width=1.2\textwidth]{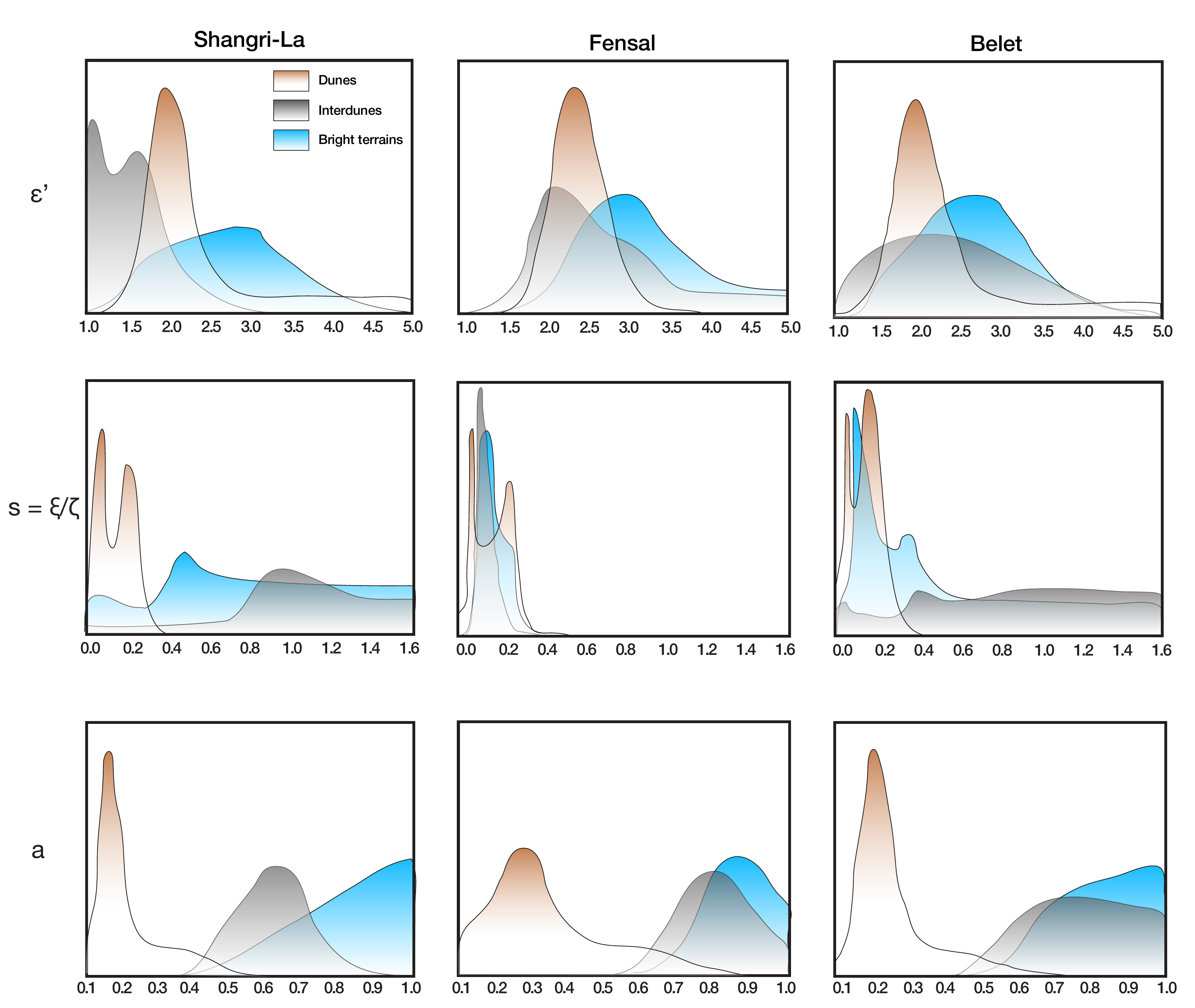}
	\caption{Marginal posterior probabilities of the relative permittivity $\epsilon'$, the surface roughness expressed as $s$ = $\xi / \zeta$, and the albedo $a$ for the considered sand seas and the units within.  Each marginal posterior probability is computed {independently} from  Markov {C}hain Monte Carlo inversion over $\sim$50 million runs.}

	\label{fig:probmarg}
\end{figure*}

The \deleted{resulting} marginal posterior probabilities computed from our Bayesian analysis for each physical parameter are summarized in Figure \ref{fig:probmarg}. We obtain nearly Gaussian posterior probabilities for the dune units for each parameter {($\epsilon'$, $\xi / \zeta$, and $a$)} {of} the three sand seas. Overall, the resulting distributions are very similar and emphasize the homogeneity of the dune material over the three studied regions.

The {results are more subtle} in the case of interdune and bright area units. The shapes of their respective marginal posterior probabilities indicate that for all of the considered physical parameters, the interdunes and bright areas are made of more complex media, possibly with mixed contributions. These distributions also suggest an important variability of surface properties across Titan at the global scale. It is worth noting that the radar-bright areas are extracted from {only inselbergs in Belet; without interdunes this population is therefore inherently more homogenous}. {Our results suggest that the radar-bright areas in Shangri-La and Fensal are  more heterogeneous}. Note that both bound Xanadu (Figure \ref{fig:mapall}). It is unclear at this stage how this {proximity} may affect our regions of interest {by, for example, influencing sediment transport}.

For the inselbergs, the model is not capable of reproducing the observations (see "no-ampV$_{pp}$" curves {in} Figure \ref{fig:fit}).  Indeed,  the albedo reaches its saturation value of 1. Hence an amplification of the  diffuse part of the backscattering signal is required in order to fit the observations {("}ampV$_{pp}$" on Figure \ref{fig:fit}). This amplification of the diffuse part is in agreement with the findings for the radar-bright region of Xanadu from previous studies \citep{Janssen2011,Janssen2016}. Accounting for this effect by multiplying the subsurface term by a factor {$>$} 2 (i.e., 3) after \cite{Janssen2016}, we go beyond the coherent backscattering effect and hence obtain the curve "ampV$_{pp}$" {in} Figure \ref{fig:probmarg}. As shown, this correction remains to be explained by some mechanism, but it allows us to partly retrieve a Gaussian distribution of the labedo $a$.
Note that these bright areas are in many ways similar to the Xanadu region. They also show similar backscatter function values compared to Enceladus for which a few SAR observations have been made \citep{LeGall2017}. In the absence of any sediment cover, icy bedrock may present high to extreme surface roughness due to high internal friction, in particular on steady surface \citep{Kietzig2010}. Such extreme surface roughness has been shown to enhance the backscatter beyond the coherent backscatter effect in lunar regolith \citep{Campbell2012}. Another explanation would come from a strong subsurface terms as proposed by {the two layer model of} \cite{Paillou2006}. {As stated above, a two layer model is beyond the scope of this work due to the untenable number of unknowns.}

The real part of the relative permittivity ($\epsilon$')  significantly differs between the three considered geomorphic units. For the Shangri-La sand sea, the dune's distribution shape is Gaussian  with a mean value around 2. The interdune areas, however, show a non-Gaussian distribution with peaks at 1.2 and $\sim$1.7 and the bright areas are pseudo-Gaussian with a mean value around $\sim$3 with large uncertainties. By comparing to laboratory measurements \citep{Paillou2008}, the derived relative permittivities reflect a homogeneous, organic-based composition for the dunes; the relative permittivity for bright terrains is more compatible with a water-ice/ammonia mixture. Therefore, the wide bi-modal distribution of the interdunes might be explained by the presence of a mixed composition between these two end-members. Thus, when the inversion tries to fit  with this non-homogenous composition\replaced{ and failed in finding}{, it fails to find} one particular value. Note that this is consistent with the presence in the interdunes of a thin and/{or} partial sediment cover over an icy-bedrock as discussed previously in \citep{Rodriguez2014}.

Contrary to other geomorphic units, the surface roughness assessed through the ratio ($\xi / \zeta$)  shows similarity between the dune units of the three sand seas.  All three bimodal distributions have peaks at $\sim$0.05 and $\sim$0.25 which give RMS slopes of 0.07 and 0.35 respectively (where RMS slope is equal to $\sqrt{2}{} \xi / \zeta$). The implications of this finding are discussed in the next section.

Lastly, the inferred contribution of the volume scattering mainly depends on the microwave albedo $a$. Low values of $a$ correspond to low volumetric energy return due  to a small penetration depth (resulting from high absorptivity of the surface material) of the medium \deleted{along the depth of penetration}. High values of $a$ correspond to higher volumetric return due to lower absorptivity and/or volume heterogeneities caused by the presence of a bedrock and/or subsurface defects. Over the dunes, we find relatively low values of $a$ (0.2-0.3) indicating that they strongly attenuate the radar signal (i.e., they appear dark in amplitude data). The albedos of interdunes (and bright areas), however, contain a strong subsurface component (i.e.,  values of 0.5-1). This implies a significant bedrock contribution beneath the organic-covered terrrains, in agreement with the results from the other two parameter inversions as well as observations by  microwave radiometry \citep{LeGall2011,LeGall2012} (low resolution) and infrared spectroscopy \citep{Barnes2008,Bonnefoy2016} (resolved).

To summarize, our Bayesian inferences show {a} clear distinction between dunes, interdunes and inselbergs in terms of roughness and composition. Ourresults provide more evidence for the global homogeneity of the dune units
and show that interdunes and  bright areas are composed of more complex and varying {types} of terrains, with {possibly} strong regional variations \explain{how can the findings be "possibly strong"? It seems like it should be one or the other.}. {I}nterdunes {systematically} present  a wider surface roughness distribution (with higher roughness), a widely distributed permittivity  and a stronger subsurface contribution than the dunes. We discuss what this implies for the aeolian system morphodynamics in the following section.

\section{Discussion}

We are approaching the problem with both expertise in microwave modeling and aeolian sediment transport. Hence, we discuss here the connections between results of the MCMC and interpretation in terms of aeolian morphodynamics.

\subsection{Interpretations on geomorphic units properties}

As shown dunes in Fensal have slightly higher relative permittivity and albedo than in Belet and in Shangri-La. Note that the incidence angle coverage goes at lower values (i.e., almost 5$^{\circ}$) for Fensal compared to the other regions. According to our sensitivity analysis, the roughness contribution is particularly strong in this incidence domain. So, one explanation would be that this is the only region where the model can very well constrain the roughness. However, this difference may also reflect actual variations due to latitudinal and altimetric controls of Titan's dune field morphometry  \citep{LeGall2012}. Fensal is indeed at both higher latitude and altitude than Belet and Shangri-La. Sediment availability is expected to decrease when latitude and altitude increase. The relative permittivity inversions show that i) dunes are compatible with a very homogeneous sand made of porous organics, ii) bright terrains  are a mixture of contaminated water-ice with tholins in different ratios, and iii) the backscatter function of the interdunes may reflect spatial variations with contributions from both organics and icy bedrock. Our results thus provide a direct confirmation of what was previously suggested from infrared analysis \citep{Barnes2008,Soderblom2007} and  passive microwave observations \citep{Janssen2016,LeGall2011} compared to laboratory characterization measurements \citep{Paillou2008}. Our derived values for the roughness  also provide  interesting insights.  The  distributions have a systematic best-fit with observations despite differences in data coverage (Figure \ref{fig:sigma0_all}).

 While we cannot directly link the actual distribution of the roughness with  our results over the dunes, compared to other units, reflect a  clear non-homogeneous distribution. Note that  the synthetic tests  emphasize that both model and method are stable even for spatially distributed values over the three main parameters.  Moreover, dunes on Earth and on Mars, are represented by smooth terrains (at micro-wave wavelengths) with secondary patterns, such as ripples and avalanches on the lee side.  Therefore,  we expect that these secondary patterns that should be present on Titan as well -- otherwise these features would not be dunes.   The explanation for non-Gaussian distribution after MCMC for the dunes is here presented as the ability of the method to detect the systematic effect (i.e., for the three regions) of such patterns in the roughness component of the microwave backscatter function over Titans' dunes. Note that such distributions are not consistent with the other, non-dune units across regions of interest. Because the  distributions we obtained show a systematic pattern across the three regions of interest (again accounting for differences in the observations described above),  we suggest that the patterns may reflect distinct surface roughness contributions. The lowest surface roughness contribution would correspond to a smooth surface where vertical roughness is one order of magnitude smaller than horizontal. The highest would be  compatible with second order bedforms of higher aspect ratio, such as ripples and/or avalanches. Uncertainties span over the obtained distributions and so do not allow us to be more quantitative on the actual roughness.

Interdunes and bright-areas exhibit wider surface roughness distributions for Shangri-La and Belet. This may be due to multi-scale roughness variations compatible with either mountainous/dissected terrains or sedimentary plains made of a non-homogeneous and widely distributed granularity, with significant contributions from higher scale roughnesses. This distinct difference  between the roughness of the dunes and the interdunes is a critical result.  It might correspond to differences in sediment size and/or variations of thicknesses of the organic cover in the interdunes, in agreement with recent analysis in the IR spectra from VIMS instrument \citep{Bonnefoy2016}. Fensal roughness distributions are, on the other hand, much narrower with close mean values. This may be indicative that Fensal's dunes, interdunes and inselbergs differ in composition (as at Shangri-La and Belet) but are similar in terms of texture (which is not the case at the other two dune fields).

To summarize our  interpretations, the dunes are organic-rich smooth terrain, potentially with detected  secondary bedforms, while interdunes are water-ice rich rough bedrock partly covered by a shallow layer of organic sediments.

\subsection{Implications on morphodynamics}

 All studies of dune morphologies and interactions with topography agree that Titan's linear and barchanoid dunes propagate eastward \citep{Lorenz2006,Radebaugh2008, Rodriguez2014,Lucas2014a,McDonald2016}. Due to our limited knowledge on the current climatic conditions, only recent Global Circulation Models (GCM) have succeeded in generating surface winds with eastward mean direction and sufficient strength to allow sediment transport \citep{Lebonnois2012,Tokano2010,Ewing2015,Charnay2015, McDonald2016}. Globally averaged surface winds on Titan are predicted by GCMs to be primarily bi-modal, blowing seasonally to the southwest or the northwest with an average speed rarely exceeding $\sim$1~m/s near the surface \citep{Tokano2010,Lebonnois2012,McDonald2016}.
Consequently, while it is commonly accepted that these aeolian systems are mature and still currently active with a ~50, 000-year timescale for Titan's dunes to reorient following global climatic changes \citep{Ewing2015}, there is a long-standing debate on whether the dune orientations reflect past or current wind conditions at Titan's surface \citep{Ewing2015,Lucas2014a,Charnay2015,McDonald2016}. The current wind conditions are \textit{a priori} incompatible with observed dune migration to the east.

In order to reconcile geomorphic observations  with wind conditions, two opposing dune evolution models have been proposed: i) reorientation of dune crests resulting from variations in sediment availability,  wind direction and strength on long-term climate cycles controlled by variations of Saturn's eccentricity (Croll-Milankovitch cycles) \citep{LeGall2012,Ewing2015,McDonald2016} and ii) equinoctial gusts and/or storms such as those observed in 2010 \citep{Turtle2011} generating winds that drive strong eastward sediment fluxes under the current conditions \citep{Tokano2010,Lucas2014a,Charnay2015}. Note that the second hypothesis succeeds in accurately explaining the dune orientation, direction of migration, as well as the equatorial dune confinement \citep{Lucas2014a}. In either case, aeolian transport is clearly the dominant transport mechanism of sediment in the equatorial zones \citep{Malaska2016}. Isolated dunes evolve by recycling their own sediment at the crest.  Hence, their sedimentary structure reflects the wind regime responsible for their formation and evolution. However, this effect is limited to a short period of time. At the scale of a vast sand sea, morphologies integrate climatic cycles punctuated by alternations between arid and humid periods.

Therefore, it is challenging to determine if a sand sea is in equilibrium with contemporary wind regimes or whether it is in a transient state reflecting different wind regimes integrated over time \citep{Ewing2015, Lucas2014a, McDonald2016}. When the sediment bed is a thick layer of particles that can be entrained by the wind, dune growth can be described as a bed instability selecting the alignment for which the normal-to-crest component of transport is maximized
\citep{RubinHunter87,Ping2014,Gadal2018}. It gives rise to the emergence of a regular dune pattern where topographic
highs (crests) and lows (troughs) keep the same composition and texture. In contrast to this transport-limited
condition, recent  works have shown that dunes may also elongate on a non-erodible surface in the direction of
the resultant sediment flux at the crest \citep{Reffet2010,Courrech2014,Gao2015,Lucas2015}. These finger-like
structures occur under multidirectional wind regimes in zones of low sand availability or where coarse grains form
an armor layer as a result of size-segregation effects (e.g., selective transport and/or avalanche as described in
\citet{Gao2016}). In all these cases, dunes composed of mobile particles can be easily distinguished from the
interdune areas (bedrock, desert pavement). The sediment reservoir  therefore critically controls  the dune
growth mechanism and  the subsequent dune and surface properties. In regards to these considerations, we hence
propose that our results on the shallow surface of
Titan's sand seas (dune and interdune areas) can  provide new insight on the local dune morphodynamics.

      \begin{figure*}[!ht]
    	\centering
    	\includegraphics[width=0.78\textwidth]{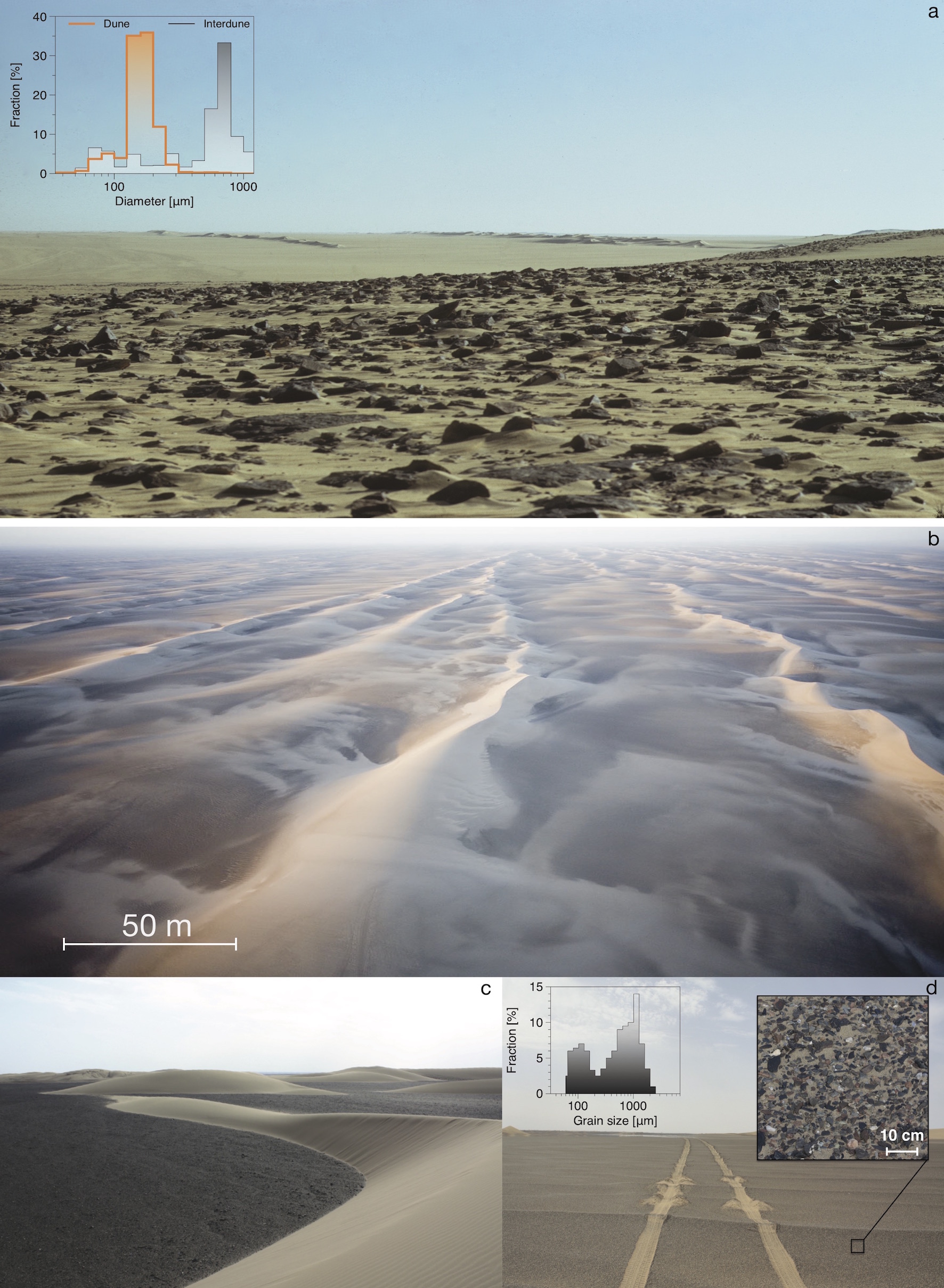}
    	\caption{Examples of size segregation in terrestrial dune fields:
		(a) T\'en\'er\'e desert (Niger);
		(b-d) Kumtagh desert (China).
		In both dune fields, the surface grain size in interdune areas is 5 to 20 times larger than on
		dunes (insets).
		As a result, the surface roughness is 1 to 2 orders of magnitude higher in the interdune areas.
		Granulometry comes from \cite{Lucas2015} and \cite{Qian2015} for Niger and China, respectively.
		Photograph credits: Y.~Callot (Niger, 1976),  G.~Steinmetz and C.~Narteau (China, 2011).}
    	\label{fig:earth}
    \end{figure*}

On Earth, dunes elongating over tens of kilometers are the prominent geomorphic features in the T\'en\'er\'e desert in
Niger \citep{Lucas2015} and in the Kumtagh desert in China \citep{Ping2017}, where they form isolated linear ridges
aligned with the resultant transport direction (Figure~\ref{fig:earth}). These longitudinal dunes are regularly
indented by transverse secondary bedforms, which may locally eject barchans. \citet{Ping2017} have shown that
non-symmetric multidirectional wind regimes like those currently measured in these deserts produce the observed dune
patterns. Indeed, the coexistence of the two dune growth mechanisms can naturally explain how finger-like structures
elongate in a specific direction and generate superimposed bedforms with a different orientation as the bed
instability develops on their flanks. The instability is not observed in the interdune areas because the bed surfaces
exhibit armor layers composed of coarse-grained material and zibars (Figure~\ref{fig:earth}).
The transport rate of uniform fine grains along the crest can then be derived from the local wind data to explain the
elongation rate of dunes as well as the migration rate of secondary bedforms \citep{Lucas2015,Ping2017}.

Although there is no established link between the grain size and the surface roughness at the RADAR wavelength, our
infrared and microwave analyses suggest that the surface layers of Titan's dune and interdune areas are different in
composition and texture.
On many (if not all)  dune fields on Earth, such differences between dunes and interdunes is the result of granular sorting causing armoring effect in aeolian systems.  A grain size ratio of 5 to 10 and a surface
roughness ratio of 10 between the interdune areas and the dunes are typically observed (Figure~\ref{fig:earth}).  Accounting for the above considerations and as we find a significantly differences in the RMS slopes, we propose that the sediment beds in the interdune areas have a mean
grain-size larger than on the dune bodies.  In Titan's equatorial belt, dunes essentially
composed of fine sediment may thus have developed over a coarser/rougher sediment layer as a result of a size-segregation process.

Such an interpretation of the compositional and textural properties of Titan's dune fields supports the elongating growth
mechanism proposed by \cite{Lucas2014a}. Considering this scenario, sediment transport occurs mainly along the linear
dunes and not in the interdune areas covered with a pavement of coarse particles. The dynamics of the size-segregation
process could also explain the origin of the dune fields in the absence of apparent source of sediment in Titan's
equatorial belt. Indeed, the mobile particles that compose the active dunes may be stored under the armor bed and
occasionally set in motion by the strong eastward winds associated with equinoctial storms \citep{Tokano2010,Charnay2015}.
In addition, Titan's dune "defects" depicted in \cite{Ewing2015} as evidence of long-term dune re-orientation from ancient
to modern winds are in fact similar to the secondary patterns of raked linear dunes documented in the Kumtagh desert
(Figures~\ref{fig:earth} and \citet{Ping2017}). These smaller scale dune features could then be the natural byproduct of
Titan's dune morphodynamics under the current wind regime. In addition to the other geomorphic processes preventing transport
\citep{Rubi09,Harp17}, the granular segregation and/or the armoring effects might contribute to the morphodynamics of
the linear dunes on Titan similarly to terrestrial cases \citep{Gao2016}.

\section{Conclusions}

Microwave backscatter functions have been derived from Cassini SAR data, processed (notably corrected from the speckle noise) and analyzed over the three major sand seas in the equatorial belt of Titan. First, we show that the backscatter function from the interdunes is clearly distinguishable from  that of the dunes. This demonstrates a clear difference in composition,  attenuation, and surface roughness. Our microwave analysis is fully consistent with hyperspectral observations from the VIMS instrument. As demonstrated by a Markov chain Monte Carlo inversion,  interdunes have a higher surface roughness, a distinct relative permittivity and a stronger subsurface contribution than the dunes. These properties may reflect a thin and/or partial sediment bed  of a mixture of fine and coarser grains over an icy crust interspersed with organics. Our results may confirm that the dunes are primary composed of organics and very homogeneous at the global scale. The bimodal surface roughness distribution we derived for the dunes is also compatible with the presence of secondary bedforms, such as ripples or avalanches.

The derived physical properties of the three major Titanian sand seas suggest that interdunes may act as an armored surface composed of large grains over which the linear dunes may develop. As coarser grains would need higher shear stresses to be moved, this implies that dunes and interdunes significantly differ in their respective contribution to the total sediment flux: the sand of the dunes largely dominates the flux. Such a scenario would be similar to what is observed in Niger or China where linear dunes elongate along the resultant drift direction under a multi-directional wind regime over a sedimentary bed dominated by pebbles and boulders which are rarely  mobilized by the winds. The resulting morphodynamics is therefore compatible with the elongating growth mechanism  proposed by \cite{Lucas2014a}.

Indeed, unless the current wind regime at Titan's surface is bi-modal and purely symmetric -- which is unlikely due to Titan's non-circular orbit, topography, and albedo variations -- one can expect a different orientation of the dune crests based on the sediment supply of the interdunes: (1) depleted and/or coarser-grained interdunes, or (2) densely covered  interdunes with the same granularity of the dunes. Furthermore, the state of the interdunes contributes to  linear dunes growth as observed in the Rub' al Khali and  in the northern hemisphere of Mars. Additionally, the different dune shapes observed on Titan (e.g., raked linear dunes, star dunes, asymmetric barchans, see \cite{Radebaugh2013,Lucas2014a,Ewing2015}) clearly suggest the presence of a complex wind regime associated with a limited sediment supply at Titan's equatorial region; these observations are also compatible with our results.

Consequently, sand seas on Titan appear to be currently evolving  under a complex wind regime, with dune pattern coarsening and interdune armoring similar to their terrestrial analogues either in Saudi Arabia,  Niger or China. Future investigation, further investigation into the orientations of potential secondary bedforms (i.e., ripples or raked-shaped linear dunes) and their sub-sequent secondary winds must account for local  azimuthal variation of the backscatter function and local slope effects will be necessary.
The methods presented here could also allow us to quantitatively retrieve
the surface properties of different geomorphic units (e.g. plains, badlands, mountainous, fluvial, and lacustrine terrains) with the highest spatial resolution possible with Cassini data. Such results would enable further investigation into Titan's geological and climatic {processes }and {inform} future in situ missions to Titan such as Dragonfly, {which is target to explore these} equatorial regions \citep{Turtle2017}. {Based on our results, the Shangri-la sand sea offers the opportunity to explore dune and interdune areas of distinct composition and texture, thereby providing Dragonfly a representative sampling of Titan's diverse sedimentary and geomorphic transport processes.

\begin{acknowledgments}
%\section*{Acknowledgments and Data}

No author declares conflicts of interests. Cassini SAR data are available on the PDS (\url{https://pds-imaging.jpl.nasa.gov/portal/cassini_mission.html}). Due to the large file size, NLDSAR data set is available upon request by contacting the corresponding author. Part of the dataset is accessible from Zenodo doi:10.5281/zenodo.528545.
The authors thank Jani Radebaugh and Catherine Neish as well as anonymous reviewers and the JGR-Planets associate editor for their constructive evaluation that helped to substantially improve the manuscript. Authors also thank St\'ephane Jacquemoud, Sebastien Labarre, L\'eon Sylvain and Emma Ay\c coberry for their fruitful inputs that improved this work. Finally, the authors also thank the Cassini Radar Science Team for constructive feedbacks during this research.
The authors thank Y. Callot and G. Steinmetz for providing field pictures from Niger and China, respectively. The authors acknowledge the financial support of the UnivEarthS Labex program at Sorbonne Paris Cit\'e (ANR-10-LABX-0023 and ANR-11-IDEX-0005-02), the ANR EXODUNES, Space Campus from Universit\'e Paris Diderot, the Nvidia Academic Research Support program as well as the CNES and Institut Universitaire de France. SMM was supported by NASA Outer Planets Research Program NNX14AR23G/118460.

\end{acknowledgments}

\section*{Appendix}
\renewcommand\thetable{A.\arabic{table}}
\renewcommand\thefigure{A.\arabic{figure}}
\setcounter{figure}{0}

\subsection*{A1. Cassini SAR data summary}
The overall incidence coverage for each respective swath used in this work is shown in Fig. \ref{fig:cov}.

\begin{figure}[!ht]
	\centering
		\includegraphics[width=.65\columnwidth]{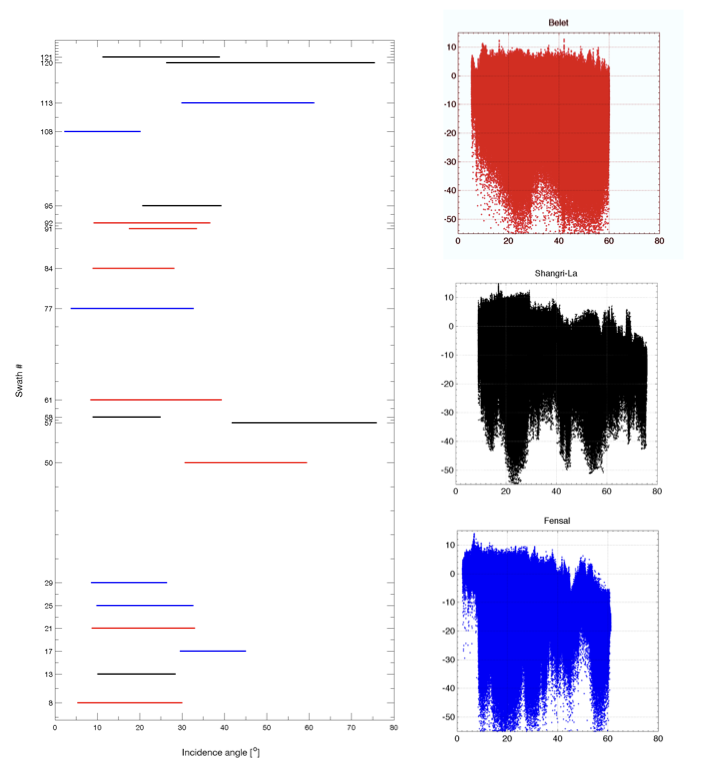}

\caption{Summary of used Cassini SAR swaths used  and their respective  incidence coverage over Belet in red, Shangri-La in black and Fensal in blue.}
\label{tab:data}
	\label{fig:cov}
\end{figure}

\subsection*{A2.  Microwave backscatter models and their domain of validity}

From very smooth to very rough terrain, 1-layer Integral Equation Method (IEM), PO (Physical Optics)  and GO (Geometrical Optics) surface scattering models are appropriate and read:

\begin{equation}
\sigma^{0}_{IEM} = \frac{k \exp(-2 k^2 \xi^2 \cos^2(\theta)) }{4}  \sum \limits_{n=1}^{\infty} |I^n_{hh} |^2 \frac{W^{n} (-2k\sin(\theta),0)}{n!},
\label{eq:ie}
\end{equation}
where $k$ is the wave number, and $W^{n}$  the Fourier transform of the nth power of the surface correlation function, which depends on $\zeta$, and with
\begin{eqnarray}
I^n_{hh}  =  & (2k\cos(\theta)\xi)^n f_{hh}  e^{-k^2 \cos^2(\theta)\xi^2} +  \nonumber \\
 &  \frac{1}{2}(k \cos(\theta)\xi)^n \left[ F_{hh} (-k \sin(\theta)) + F_{hh}(k \sin(\theta)) \right],
\end{eqnarray}
with $f_{hh}$ and $F_{hh}$ being the coefficients of the Kirchhoff and complementary fields \citep{Fung1992}{;} and
\begin{eqnarray}
\sigma^{0}_{POM}  = &  2 k^2 \cos^2 (\theta)   \Gamma_{hh} (\theta) \exp(-2 k \xi \cos(\theta))^2 \times \nonumber \\
 & \sum \limits_{n=1}^{\infty}  \frac{(2 k \xi \cos (\theta))^{2n}}{n!} \int_0^{\infty} \rho^n(x) J_0 (2k x \sin(\theta)) x dx,
\label{eq:po}
\end{eqnarray}
with $J_0$, $\Gamma_{hh}$ and $\rho(x) = \exp(-x^2 /\zeta^2) $ being respectively the zeroth-order Bessel function of the first kind, the Fresnel reflectivity and the Gaussian surface autocorrelation function{;}  and
\begin{equation}
\sigma^{0}_{GOM} = \Gamma(\theta = 0) \frac{\exp(-\tan^2 (\theta) /2m^2)}{2m^2 \cos^4 (\theta)},
\label{eq:go2}
\end{equation}

with $m = \sqrt 2 \xi / \zeta$ and $\Gamma(\theta = 0)$ being the Fresnel reflectivity at normal incidence. Note that the coherent part of the backscatter is correctly simulated with the GO model which is  therefore valid at all incidences.  The non-coherent surface scattering model is selected as a function of the roughness domain relative to the RADAR wavelength.  From smooth to  very rough terrain,  IEM,  POM and GOM are alternatively considered for the modeling of the surface backscattering coefficient $\sigma^0_{Shh}$ \citep[e.g.,][]{Paillou2006}:

\begin{equation}
\sigma^0_{Shh} = \left\lbrace
\begin{array}{ll}
\sigma^{0}_{IEM} &   \textrm{if }{} k\xi <3 \textrm{ and }{} k^2\xi \zeta < 1.5 \sqrt{\epsilon}, \\
\sigma^{0}_{POM} & \textrm{if }{} k\zeta >6 \textrm{ and }{} k^{-1} < \xi < 0.06 k \zeta^2, \\
\sigma^{0}_{GOM} & \textrm{if }{} (2k\xi \cos\theta)^2 > 10 \textrm{ and } k\zeta > 6.
\end{array}
\right.
\label{eq:sppchoice}
\end{equation}

We checked that the three $\sigma^0_{Shh}$ models provide the continuity of the solutions at the limits of each validity domain (Figure \ref{fig:valid}). We run 5.10$^5$ Monte Carlo simulations accounting for 1-layer models with no model parity and found that most (>90\%) of our reduced $\chi^2$-tests best fits  are obtained with the GOM for the 3 geomorphic units considered.  We hence obtain realistic synthetic data that mimic the actual data from Cassini SAR (Figure \ref{fig:dirflo}).  Note that this is compatible with previous works \citep{Paillou2014}.

  \begin{figure}[!ht]
  	\centering
  	\includegraphics[width=0.95\textwidth]{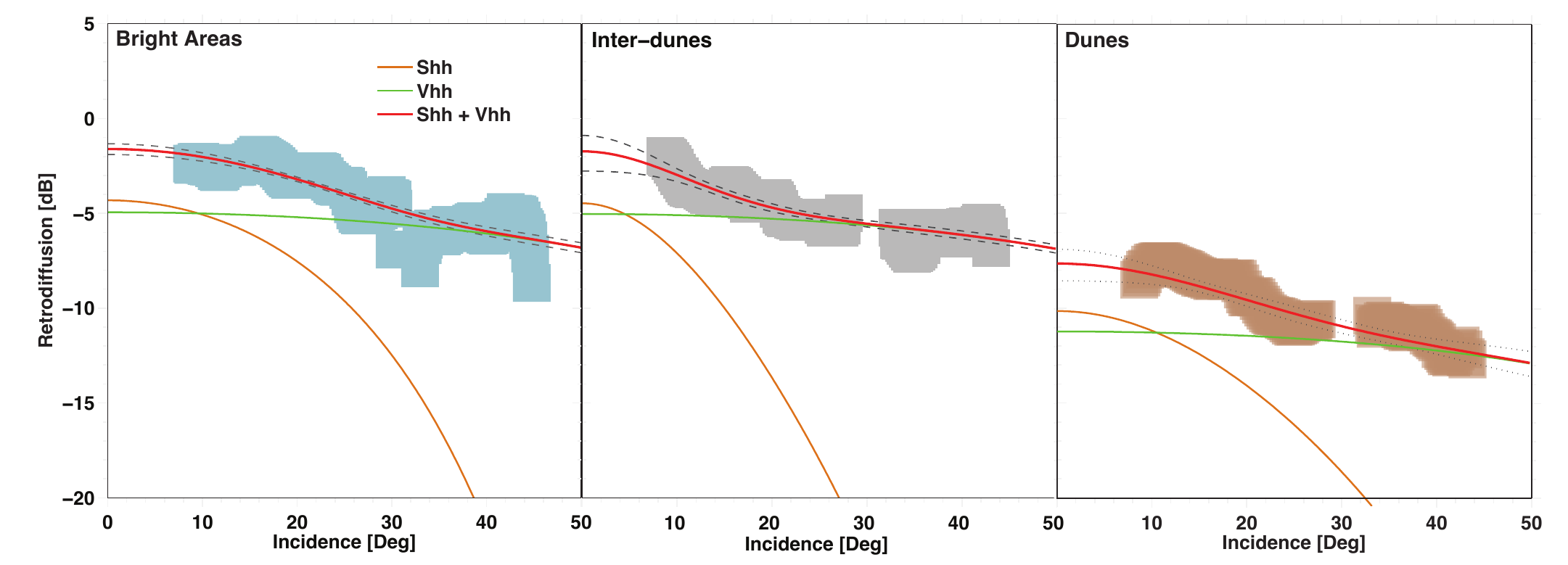}
  	\caption{Results of the direct Monte Carlo simulation obtained from the GO model over the sand sea of Belet. The colorized areas correspond to the observed backscatter signal at the three different geomorphic units with a 1-$\sigma$ standard deviation in width. The contribution of the two terms  for the surface (S$_{hh}$) and volume (V$_{hh}$) are shown separately and additively as indicated. The area embraced by the dashed-lines correspond to the best 5\% results from the $\chi^2$-tests.}
  	\label{fig:dirflo}
  \end{figure}

\begin{figure}[!ht]
	\centering
	\includegraphics[width=.55\columnwidth]{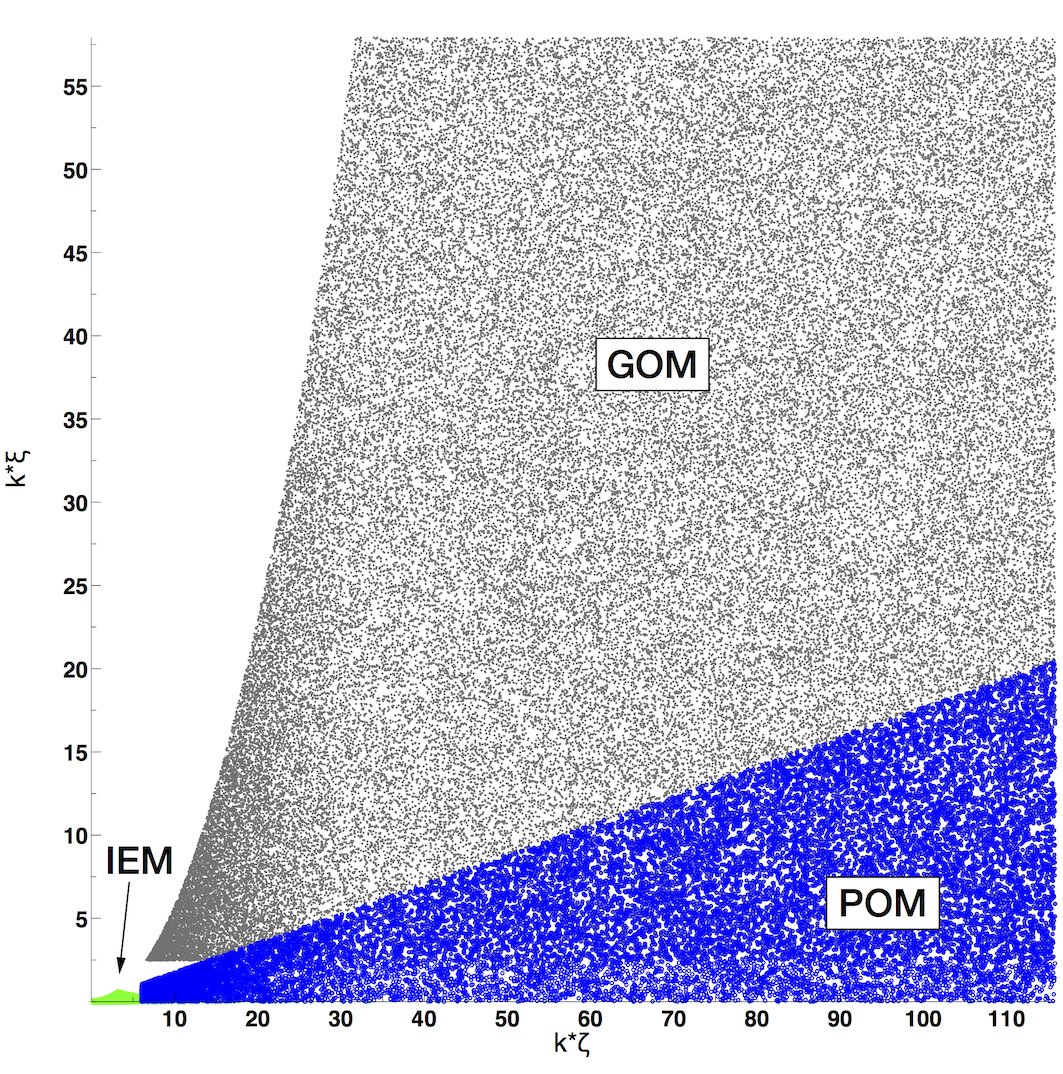}
	\caption{Domain of validity for each backscatter models}
	\label{fig:valid}
\end{figure}

\begin{figure}[!ht]
	\centering
	\includegraphics[width=.95\columnwidth]{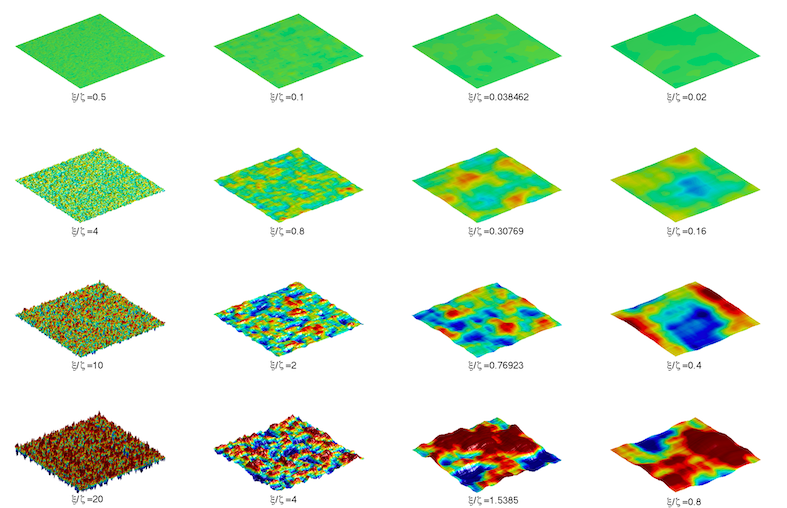}
	\caption{Generated surface of 1m$^2$ for considered $\xi$ = \{0.5, 4 ,10,20\} cm (from left to right) and $\zeta$ = \{1, 5 , 13, 25\} cm (from top to bottom). }
	\label{fig:surf_stats}
\end{figure}

\subsection*{A3. Surface roughness of natural terrains}

As discussed in the main text, statistical descriptors of natural surface roughness depends on the scale and the nature of the considered terrain. Here are combined RMS height ($\xi$) and correlation length ($\zeta$) from literature and photogrammetric surveys on agriculture terrains performed by the corresponding author (Figure \ref{fig:rough}).  These terrains exemplify the range of surface roughness observed in nature and are to be compared with respect to the values retrieved in this study. Note that there is no direct link between the nature of the terrain and its surface roughness as the latter is controlled by different processes that sculpt the topography. Also, multi-scale effects of the surface roughness are known to play a significant role in  the radiometry \citep{Labarre2017}.  Further investigation into such multi-scale effects is necessary but beyond the scope of this paper.

\begin{figure}[!ht]
	\centering
	\includegraphics[width=.45\columnwidth]{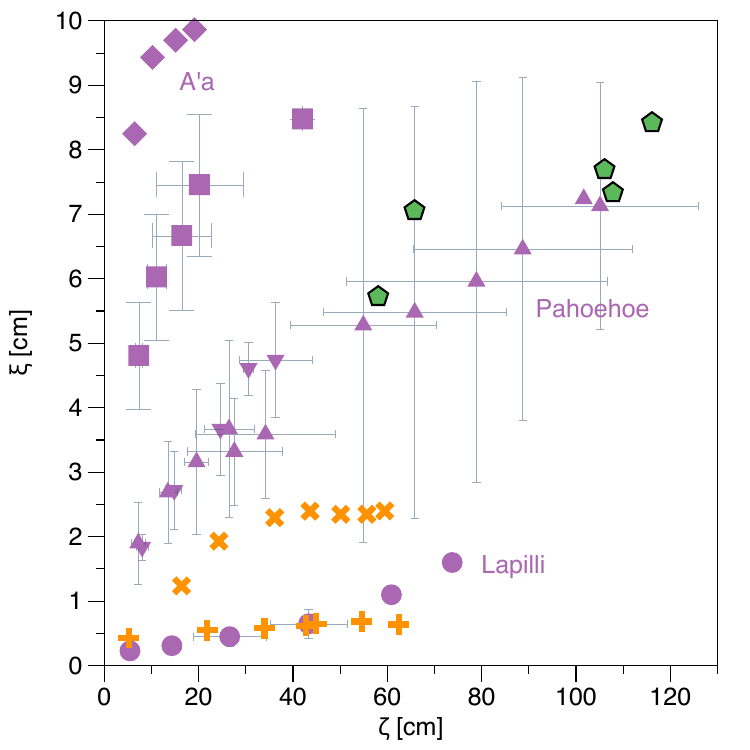}

		\includegraphics[width=.45\columnwidth]{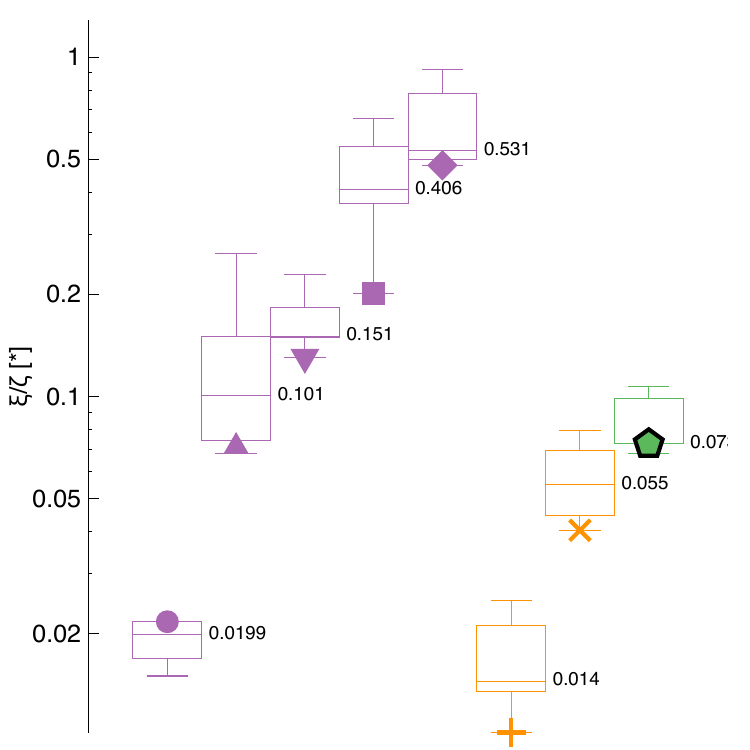}
	\caption{Natural surface roughness statistics for argiculture (green), sedimentary (orange crosses) and volcanic terrains (purple plain symbols) from lapilli to a'a lava flows. Data are collected from \cite{Baghdadi2000,Bretar2013} and UAV photogrammetric survey  over agriculture terrain performed in October 2016 by the corresponding author.}
	\label{fig:rough}
\end{figure}

\subsection*{A4. Synthetic tests on surface physical properties and subsequent microwave backscatter cross-section}

Synthetic tests are done in order to show the effect of spatial variations of the physical properties that control the microwave backscatter signal ($\sigma^0$). Synthetic maps are generated (Figure \ref{fig:synthmap}) and the resulting $\sigma^0$. Note that shadowing and layer over effects are not considered when synthesizing the SAR image. Multiplicative speckle noise with a gamma distribution has been considered based on noise estimation performed in \cite{Lucas2014b}.

	\begin{figure}[!ht]
	\centering
	\includegraphics[width=.7\columnwidth]{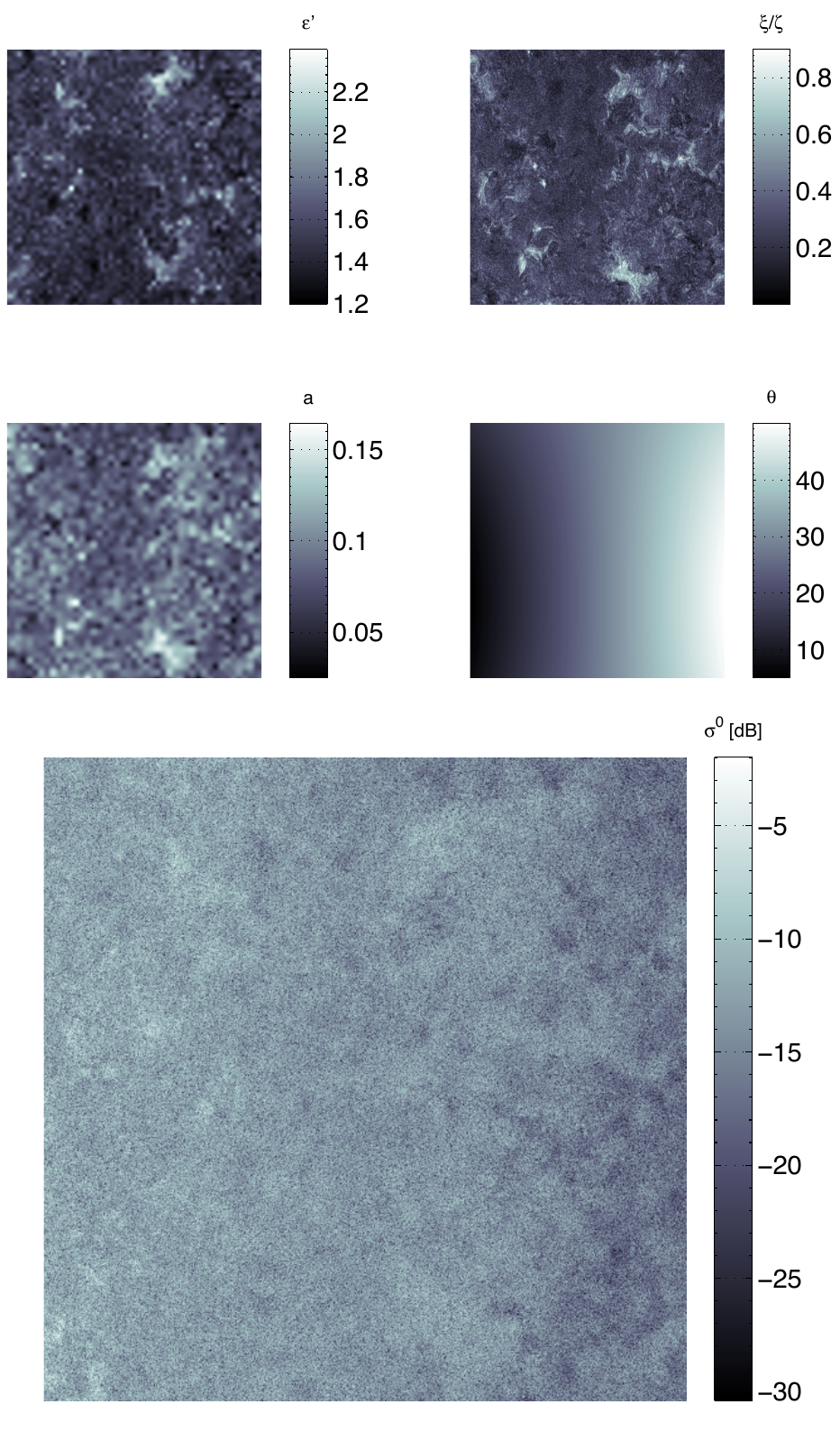}
	\caption{Example of synthetic microwave backscatter cross-section ($\sigma^0$) from physical parameters: The real part of the relative permittivity $\epsilon '$, the surface slope $\xi / \zeta$, the radar albedo $a$ and incidence angle $\theta$. }
	\label{fig:synthmap}
	\end{figure}

	\begin{figure}[!ht]
	\centering
	\includegraphics[width=.7\columnwidth]{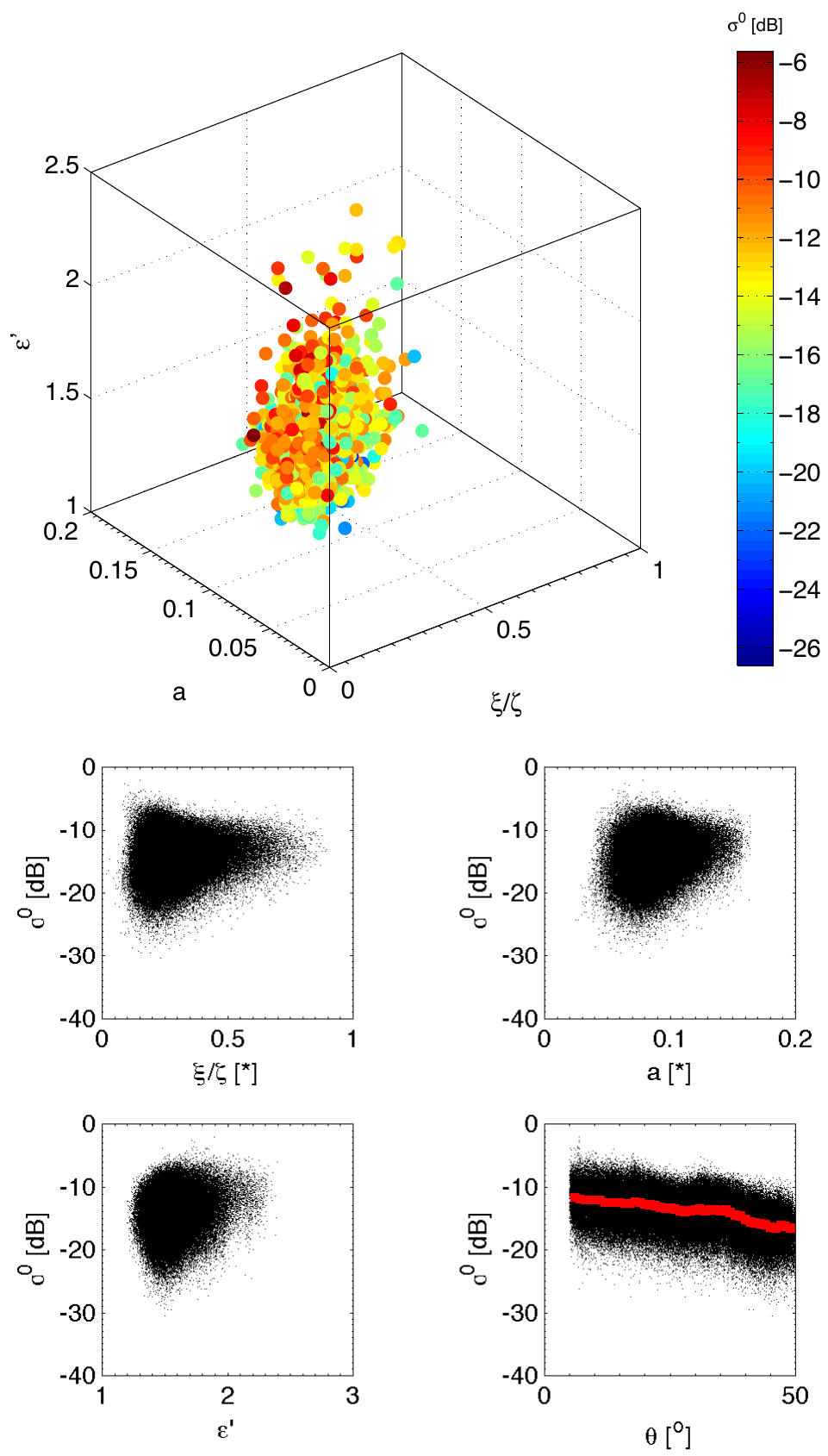}
	\caption{Respective dependencies of $\epsilon '$, $\xi / \zeta$, and $a$ on the resulting $\sigma^0$.   }
	\label{fig:synth_syn}
	\end{figure}

	\begin{figure}[!ht]
	\centering
	\includegraphics[width=.69\columnwidth]{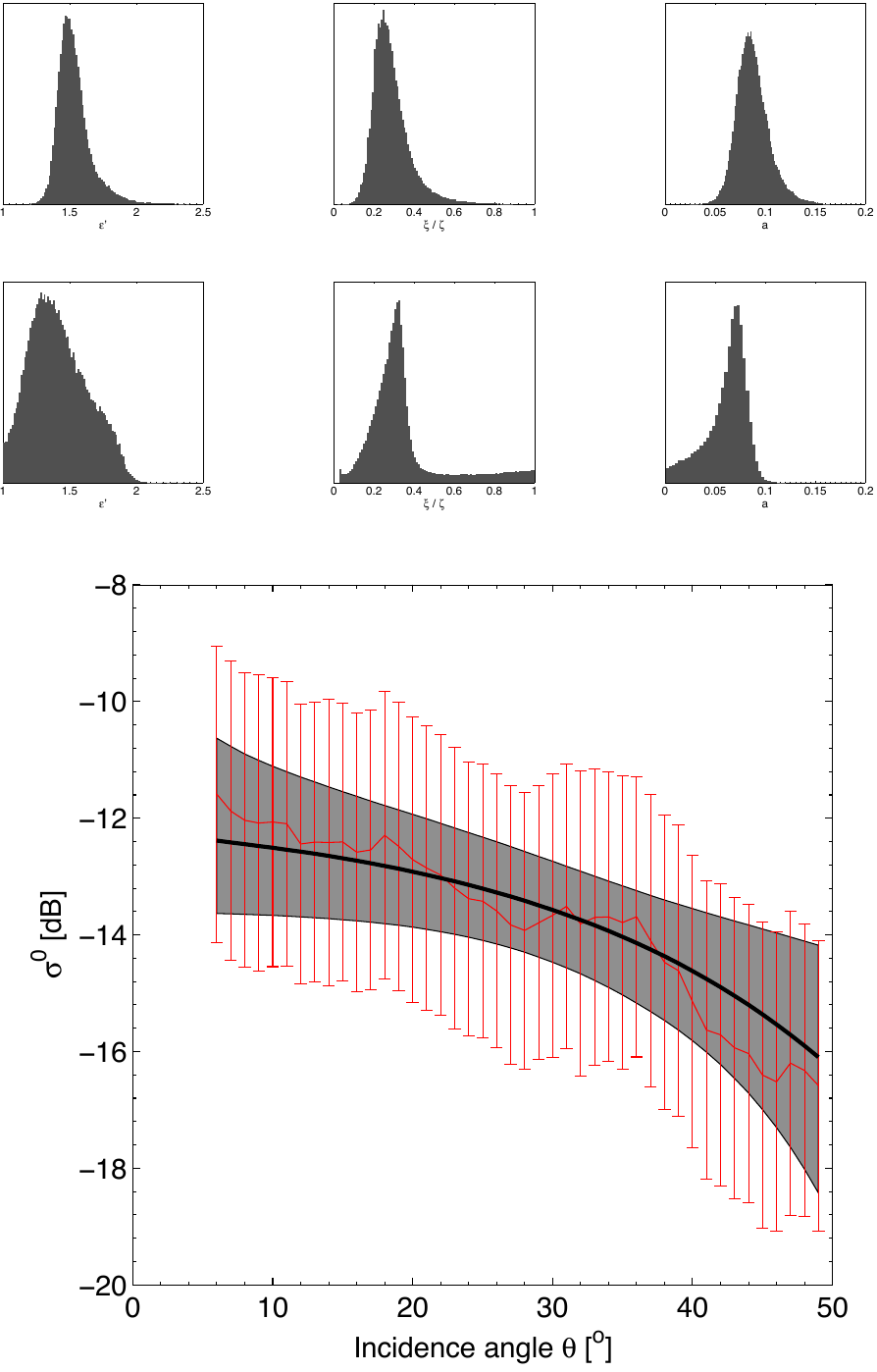}

	\caption{Respective distribution of $\epsilon '$, $\xi / \zeta$, and $a$.}
	\label{fig:synthist}
\end{figure}

\renewcommand\thetable{}
\renewcommand\thefigure{}
\setcounter{figure}{0}

\listofchanges
\end{document}